\documentclass[lettersize,journal]{IEEEtran}
\usepackage{amsmath,amsfonts,amssymb,graphicx}
\usepackage{amsmath, bm}
\usepackage{algorithmic}
\usepackage{graphicx}
\usepackage{textcomp}

\usepackage{enumitem}
\usepackage{xcolor}
\usepackage{subcaption}
\usepackage{algorithm,algorithmic}
\usepackage{relsize}
\usepackage[export]{adjustbox}
\usepackage{array,multirow}
\usepackage{multirow}
\usepackage{lscape}
\usepackage{bbm}
\usepackage{graphicx}
\usepackage{booktabs}
\usepackage{url}

\DeclareMathOperator{\SDE}{PCDC}
\DeclareMathOperator{\MoDL}{MoDL}
\DeclareMathOperator{\mmd}{MMD}
\DeclareMathOperator{\DP}{DP}
\newenvironment{mytheorem}[1][]
{%
    \par\noindent\textbf{Theorem 1.} \ignorespaces
}

\begin{document}

\title{Robust Physics-based Deep MRI Reconstruction Via Diffusion Purification}




\author{Ismail Alkhouri$^*$, \textit{Member, IEEE}, Shijun Liang$^*$, \textit{Member, IEEE}, Rongrong Wang, Qing Qu, \textit{Member, IEEE}, Saiprasad Ravishankar, \textit{Senior Member, IEEE}

\thanks{*~Equal contribution. This work was supported in part by the National Science Foundation (NSF) under grants CCF-2212065 and CCF-2212066.
}
\thanks{I. Alkhouri is with the Electrical Engineering \& Computer Science (EECS) Department, University of Michigan, Ann Arbor, MI, 48901, and with the Computational Mathematics, Science and Engineering (CMSE) Department, Michigan State University, East Lansing, MI 48824 (ismailal@umich.edu;alkhour3@msu.edu)}
\thanks{S. Liang is with the CMSE Department, Michigan State University, East Lansing, MI 48824 (liangs16@msu.edu).}
\thanks{R. Wang is with the CMSE and Mathematics Departments, Michigan State University, East Lansing, MI 48824 (wangron6@msu.edu).}
\thanks{Q. Qu is with the EECS Department, University of Michigan, Ann Arbor, MI, 48901 (qingqu@umich.edu).}
\thanks{S. Ravishankar is with the CMSE Department and the Department of Biomedical Engineering, Michigan State University, East Lansing, MI 48824 (ravisha3@msu.edu).}
\thanks{© 2025 IEEE. Personal use of this material is permitted. Permission
from IEEE must be obtained for all other uses, in any current or future media, including reprinting/republishing this material for advertising or promotional purposes, creating new collective works, for resale or redistribution to servers or lists, or reuse of any copyrighted component of this work in other works.}
}

\maketitle

\begin{abstract}
Deep learning (DL) \textcolor{black}{supervised} techniques have been extensively employed in magnetic resonance imaging (MRI) reconstruction, delivering notable performance enhancements over traditional non-DL methods. \textcolor{black}{Nonetheless, \textcolor{black}{these models have vulnerabilities} during testing such as their susceptibility to 
worst-case or noise-based measurement perturbations, variations in training/testing settings like acceleration factors, contrast, k-space sampling locations, and distribution shifts \textcolor{black}{stemming from unseen lesions and different anatomies}.} This paper addresses \textcolor{black}{these} robustness challenges by leveraging diffusion models. In particular, we present a robustification strategy that improves the resilience of DL-based MRI reconstruction methods by utilizing pre-trained diffusion models as \textcolor{black}{purifiers.} \textcolor{black}{We dub our method as \underline{RO}bust \underline{D}L-based MR\underline{I} with Diffusion Purificati\underline{O}n (\underline{RODIO}).} 
In contrast to conventional robustification methods for DL-based MRI reconstruction, such as adversarial training (AT), our proposed approach eliminates the need to tackle a minimax optimization problem. It only necessitates \textcolor{black}{efficient} fine-tuning on purified examples. Our experimental \textcolor{black}{results} underscore the effectiveness of our approach in addressing the mentioned instabilities, outperforming standalone diffusion-based MRI reconstructors and leading robustification methods for deep \textcolor{black}{supervised} MRI reconstruction, including AT and randomized smoothing. Our experiments demonstrate: \textcolor{black}{(\textit{i}) the adaptability of our approach across multiple DL-based supervised MRI reconstruction models}; \textcolor{black}{(\textit{ii}) compatibility with accelerated diffusion-based samplers}; \textcolor{black}{(\textit{iii}) robustness to data with unseen lesions}; and \textcolor{black}{(\textit{iv}) effectiveness when applied to unsupervised single-shot generative reconstructors.}

\end{abstract}

\section{Introduction}
\label{sec:introduction}

Magnetic resonance imaging (MRI) is a widely used clinical tool for visualizing anatomical and physiological structures. However, its data acquisition is typically slow due to sequential processes. To overcome this limitation, various techniques\cite{lustig2008compressed,yang2010fast,aggarwal2018modl} have emerged, enabling precise image reconstruction from limited, rapidly acquired data.

Deep learning (DL) serves as a tool for tackling large-scale inverse problems and addressing challenges in image reconstruction~\cite{Schlemper2019Sigma-net:Reconstruction,Ravishankar2018DeepReconstruction,aggarwal2018modl,Schlemper2018AReconstruction}. This study is centered on the application of DL in the context of MRI reconstruction. Among various DL techniques, established networks designed for denoising images or sensor data are essential components. The widely adopted U-Net architecture~\cite{Unet, han2018framing, lee2018deep} has been utilized to remove artifacts in MRI data resulting from undersampling. There has been particular interest in hybrid approaches that combine neural networks with imaging physics, including forward models. A particular notable example is MoDL (Model-based reconstruction using Deep Learned priors)~\cite{aggarwal2018modl}, which employs an iterative approach to address the regularized inverse problem in MRI reconstruction.

Recent research highlights potential vulnerabilities in DL-based MRI reconstruction models \textcolor{black}{during testing}, particularly susceptibility to \textcolor{black}{small additive disturbances~\cite{antun2020instabilities,zhang2021instabilities, gilton2021deep, jia2022robustness, chan2021local}. 
Test-time variations in acquisition settings, etc., could further pose challenges for generalization of learned models, leading to reduced performance and potential diagnostic inaccuracies. In addition to these vulnerabilities, DL-based MRI reconstruction may encounter other challenges in the form of domain shifts, artifacts, and test-time pathologies and lesions~\cite{kondrateva2021domain}.
Learning robust and generalizable reconstruction models from often limited available training sets is thus an important goal.}

Various techniques have been developed to enhance the robustness of DL-based MRI reconstruction tasks~\cite{jia2022robustness,wolfmaking}. One noteworthy approach, adversarial training (AT) \cite{madry2017towards}, originally \textcolor{black}{proposed} to enhance the robustness of image classifiers~\cite{madry2017towards,zhang2019theoretically,cohen2019certified,salman2020denoised}, involves solving a computationally intensive minimax optimization problem that incorporates generating adversarial examples. Another method, randomized smoothing (RS)~\cite{salman2020denoised}, also initially designed for image classifiers, smoothens network outputs when handling inputs perturbed by random noise. Nevertheless, both AT and RS have demonstrated limitations in their performance when encountering previously unseen disturbances or dealing with larger perturbation bounds.\par

Within the domain of image classification, a recent study conducted by Nie et al.~\cite{pmlr-v162-nie22a} has introduced a robustification strategy that effectively \textcolor{black}{mitigates the impact of} additive worst-case perturbations, harnessing the power of diffusion models (DMs) \cite{chung2022score, chung2022diffusion, karras2022elucidating}. 
Drawing inspiration from this methodology \textcolor{black}{and benefiting from the \textcolor{black}{potential} generalization capabilities of DMs}, we investigate the application of a similar approach to enhance the \textcolor{black}{robustness} of DL-based 
MRI reconstruction. Our approach centers on the application of pre-trained diffusion models \textcolor{black}{as noise purifiers}. More precisely, this purification process entails a gradual introduction of noise, followed by the refinement of the noise through the utilization of the pre-trained DM. 
In what follows, we outline the contributions of this article.


\subsection{Contributions}

\begin{itemize}[left=0pt]
    \item \textcolor{black}{We introduce a general robustification framework designed to enhance the resilience of \textcolor{black}{supervised} DL-based MRI reconstructors against \textcolor{black}{a variety of instabilities, and improve their generalization performance when faced with out of distribution samples}. This is accomplished through integrating purification via pre-trained DMs into existing DL-based models.}

      \item We prove that the perturbed and clean images' distributions (and conditional distributions) get closer to each other as the time \textcolor{black}{increases in the forward diffusion stage.} 

      \item We present an approach to select a process-switching time step - a critical parameter within our DM-based purification method. This \textcolor{black}{mitigates} the necessity of treating it as a hyper-parameter.

    \item \textcolor{black}{We use fine-tuning to further improve the DL-based reconstructors' performance, which, unlike well-known \textcolor{black}{state-of-the-art} (SOTA)  robustification method AT, neither requires solving a minimax problem nor involves generating \textcolor{black}{worst-case} examples. }

    \item In our experimental results, we demonstrate the effectiveness of our proposed approach by assessing it against standard evaluation metrics, \textcolor{black}{surpassing} the performance of AT, RS, and diffusion-based MRI reconstruction in the presence of \textcolor{black}{several sources of instabilities}. Furthermore, we illustrate that after being trained on the knee fastMRI dataset, the purification process using DMs extends its benefits to other MRI datasets, including a brain MRI dataset or data with unseen lesions. Additionally, we show that \textcolor{black}{(\textit{i})} our robustification approach can be applied to multiple DL-based supervised methods such as the well-known MoDL and the recent Recurrent Variational Network (RecurrentVarNet) \cite{yiasemis2022recurrent}, \textcolor{black}{(\textit{ii}) our approach can be adapted to accelerated samplers for diffusion purification, and (\textit{iii}) our method can be used to robustify unsupervised single-shot DL-based reconstructors such as the deep image prior (DIP) based model in \cite{alkhouri2024image}.}

    
\end{itemize}

\subsection{Organization}

The organization of this paper is as follows. Section~\ref{sec: related} covers related work. Section~\ref{sec: prel} presents preliminaries and motivation. In section~\ref{sec: prop method}, we introduce our DM-based robustification approach for Deep MRI Reconstruction models. Section~\ref{sec: exp} showcases experimental results, and section~\ref{sec: conc} concludes our study.

\section{Related work}
\label{sec: related}
Various DL-based methods have been introduced for MRI reconstruction. An example is the ADMM-Net~\cite{yang2017admmnet}, which uses neural networks to determine the parameters for the ADMM algorithm. In \cite{zhang2018istanet}, the authors introduced a technique based on the Iterative Shrinkage-Thresholding Algorithm (ISTA) to optimize a general $\ell_1$ norm reconstruction model. On the other hand, MoDL \cite{aggarwal2018modl} combines model-based reconstruction with DL, utilizing a data-consistency term and a learned NN to capture image redundancy. The NN parameters are determined through end-to-end supervised learning with respect to (w.r.t.) the unrolled iterative process. A more comprehensive review on DL-based image reconstruction methods can be found in \cite{8844696, 10004819, ongie2020deep}.\par

In recent years, various robustification techniques have emerged in the field of image classification \cite{Wong2020Fast,madry2017towards,zhang2019theoretically,cohen2019certified,salman2020denoised}. These methods leverage formulations relying on either the minmax optimization of adversarial learning or randomized smoothing. Notably, these robustification strategies have been applied in deep MRI reconstruction as well \cite{jia2022robustness,wolfmaking}. In the work by Jia et al. \cite{jia2022robustness}, the authors proposed the use of AT and data augmentation to enhance the robustness of image reconstruction methods against worst-case additive perturbations and image transformations. On the other hand, in the study by Wolf et al. \cite{wolfmaking}, an end-to-end Randomized Smoothing (E2E-RS) approach was employed to enhance \textcolor{black}{DL-based MRI reconstructors} against worst-case additive perturbations. However, it's important to note that both of these methods exhibit limitations. They tend to experience performance degradation when dealing with higher perturbation \textcolor{black}{bounds and out of distribution test measurements}. Additionally, they do not address other potential instabilities inherent in the MRI reconstruction problem. Our work, focused on robustification, distinguishes itself from these two approaches by (\textit{i}) utilizing a purification pipeline based on pre-trained DMs and (\textit{ii}) addressing additional instabilities stemming from \textcolor{black}{(\textit{a})} disparities in acquisition criteria between the training and testing phases, and \textcolor{black}{(\textit{b}) changes in anatomy and presence of pathologies or lesions}. \textcolor{black}{Among other methods,} we will use AT and E2E-RS as baselines for our robustness comparisons.\par 

Diffusion models have recently attracted significant attention across a range of DL applications. Leveraging advancements in computer vision, the field of medical imaging has seen a notable rise in interest in diffusion models (such as \cite{chung2022score} and \textcolor{black}{\cite{chung2024decomposed}). We refer the reader to \cite{kazerouni2023diffusion} for a survey on diffusion-based applications in medical imaging}. 
In the context of MRI reconstructions, several diffusion-based approaches have been introduced, as highlighted in \cite{xie2022measurement, gungor2023adaptive, peng2022towards, chung2024decomposed}. 

The authors in \cite{chung2022score} introduced a DM-based approach for solving the Deep MRI reconstruction inverse problem. In their research, they propose incorporating a data consistency step into the reverse process, enabling the sampling from a conditional distribution which is a key component in many diffusion-based inverse problems solvers such as \cite{song2023solvingHard}. \textcolor{black}{Following \cite{chung2022score}, the authors in \cite{chung2024decomposed} introduced Decomposed Diffusion Sampling (DDS) that provided more efficient sampling in the reverse process.} However, our work in this article differs in two key aspects. Firstly, our primary objective is to enhance the robustness of DL-based reconstruction approaches against different perturbations, a facet not addressed in \cite{chung2022score,chung2024decomposed}. Secondly, we utilize pre-trained DMs as purifiers to eliminate artifacts and perturbations in DL-based reconstructors. Specifically, in their method, the reverse process begins with random Gaussian noise aiming to reconstruct the image, while our proposed purification method \textcolor{black}{starts} from the initial aliased image. Subsequently, we gradually add noise to some time step before starting the reverse process. It is also important to note that, while using the same sampling algorithm, the number of time steps required in our method is much lower than the those needed in \cite{chung2022score}. \textcolor{black}{The methods in \cite{chung2022score,chung2024decomposed}} will be used as another baseline in our experiments. To the best of our knowledge, our proposed method is the first attempt to robusify \textcolor{black}{supervised} DL-based MRI image reconstructors using DMs. The recent study conducted by \cite{oh2023annealed} introduced an approach to mitigate MRI motion artifacts through diffusion purification coupled with data consistency. Their method involves training a deep learning model solely on motion-free images, followed by the repeated application of forward and reverse diffusion processes to gradually enforce low-frequency data consistency. While our approach shares similarities with theirs in utilizing diffusion purification, it is essential to emphasize that our proposed method employs diffusion purification as a precursor to robustify deep learning-based supervised MRI reconstructors against various sources of instabilities, and enhance their generalization performance. Moreover, in contrast to \cite{oh2023annealed}, which necessitates multiple forward and reverse passes through the diffusion model, our method only requires one forward and backward pass. 

While the main focus of this study is the robustness of supervised methods, DL-based unsupervised models have also been explored in the context of MRI reconstruction. Some DL scan-specific methods involve training a neural network to learn how to interpolate the subsampled regions of k-space from the fully-sampled autocalibration scan region \cite{kim2019loraki, tamir2016generalized, arefeen2022scan, haldar2015autocalibrated}. In Section~\ref{sec: exp}, we evaluate the robustness of our proposed approach against the scan-specific method in \cite{kim2019loraki}.  

The study in \cite{pmlr-v162-nie22a} introduced an adversarial purification method using DMs to improve NN image classifiers' robustness, a context distinct from inverse imaging. Our work focuses on DL-based MRI reconstruction, which is both an inverse and regression problem. Both our work and \cite{pmlr-v162-nie22a} require selecting a key parameter, the process-switching time (PST) step. However, unlike \cite{pmlr-v162-nie22a}, which necessitates PST step tuning, we propose a novel PST step selection method based on the well-known Maximum Mean Discrepancy (MMD) metric \cite{gretton2006kernel}. Moreover, we note that this selection method is applicable to any DM-based purification approach.\par

It is important to clarify that when we use the terms `adversarial' and \textcolor{black}{`attacks'}, we are referring to the worst-case additive noise w.r.t. \textcolor{black}{the DL-based MRI reconstructor} and the perturbation budget \textcolor{black}{used for robustness evaluation}, as further explained in the upcoming sections.


\section{Lack of Robustness in DL-based MRI Reconstruction \& Score-based DMs}
\label{sec: prel}

In this section, we first introduce the inverse problem formulation for Deep MRI reconstruction. \textcolor{black}{Second, we shed light on} the lack of robustness in these models. Then, we present the formulation of the score-based DM used in this paper. 

\subsection{DL-based MRI Reconstruction}

MRI reconstruction is a challenging ill-posed inverse problem~\cite{compress}. Its objective is to recover the original signal $\mathbf{x} \in \mathbb{C}^{n}$ from observed measurements $\mathbf{y} \in \mathbb{C}^{m}$, with $m < n$. \textcolor{black}{For multi-coil MRI, this task can be formulated as a linear inverse problem denoted as $\mathbf{y} \approx \mathbf{A} \mathbf{x}$, where $\mathbf{A} = \mathbf{M} \mathbf{F} \mathbf{S}$ with $\mathbf{S}$ denoting the sensitivity encoding with multiple coils, $\mathbf{F}$ denoting coil-by-coil Fourier transform, and $\mathbf{M}$ denoting coil-wise undersampling}.

Typically, the reconstruction process involves solving the optimization problem $\min_{\mathbf{x}}  \{ \|\mathbf A \mathbf{x} - \mathbf{y} \|^{2}_2 + \lambda \mathcal{R}(\mathbf{x})\}\:,$
where $\mathcal{R}(\cdot)$ (resp. $\lambda > 0$) is a regularization term (resp. parameter).\par

There are several methods that use unrolling steps to train Deep MRI image reconstruction. While for the major part of this paper we focus on the popular MoDL framework \cite{aggarwal2018modl}, our proposed method can be applied to other DL-based reconstruction models, as illustrated in the last subsection of the experimental results. In MoDL, the traditional regularization term is substituted with a denoising Neural Network (NN) represented as $f : \mathbb{C}^n \rightarrow \mathbb{C}^n$, parameterized by $\theta$. This denoising NN is trained in a supervised learning framework using a dataset of multiple pairs of measurements $\mathbf{y}$ and their corresponding ground truth images $\mathbf{x}$.\par

For each pair $(\mathbf{y},\mathbf{x})$ in the training set $D$, the MoDL training process initializes $\mathbf{x}_0$ (e.g., as $\mathbf{A}^H \mathbf{y}$) and then iterates through the subsequent steps for a specified number of unrolling iterations indexed by $j\in \{0,\dots,N-1\}$. This process can be described as follows:
\begin{equation}
\label{eqn: MoDL training 1}
\mathbf{z}_j \leftarrow f_\theta(\mathbf{x}_j) \:,
\end{equation}
%
\begin{equation}
\label{eqn: MoDL training 2}
\mathbf{x}_{j+1} \leftarrow \underset{\mathbf{x}}{\arg \min} \|\mathbf{A}\mathbf{x}-\mathbf{y}\|^2 + \lambda  \|\mathbf{x}-\mathbf{z}_j\|^2  \:.
\end{equation}
The parameters of $f_\theta$ are updated end-to-end in a supervised manner following \cite{aggarwal2018modl}. 

Equation \eqref{eqn: MoDL training 1} corresponds to the denoising step, while Equation \eqref{eqn: MoDL training 2} pertains to the data consistency (DC). Equation \eqref{eqn: MoDL training 2} has a closed-form solution given by $\mathbf{x}_{j+1} \leftarrow (\mathbf{A}^H \mathbf{A} +\lambda \mathbf{I})^{-1}(\mathbf{A}^H \mathbf{y}+\lambda \mathbf{z}_j)$.

During the testing phase, when presented with an aliased image (e.g., $\mathbf{A}^H\mathbf{y}$), a trained MoDL model reconstructs $\mathbf{x}$ by applying the procedure described in Equations \eqref{eqn: MoDL training 1} and \eqref{eqn: MoDL training 2} for a specified number of unrolling steps. For the remainder of this paper, we use $\MoDL_\theta (\mathbf{A}^H \mathbf{y})$ to denote the image reconstructed from MoDL.\par


\subsection{Vulnerabilities of DL-based MRI Reconstructors}
\label{sec: Vul}

\subsubsection{\textcolor{black}{K-space Additive Noise}}

Given a trained \textcolor{black}{deep MRI} reconstruction NN and an aliased image $\mathbf{z} = \mathbf{A}^H \mathbf{y}$, recent studies have shown that \textcolor{black}{these NNs are} not robust to additive perturbations $\bm{\delta}$ to $\mathbf{y}$ \cite{li2023smug}. The study in \cite{jia2022robustness} presents an \textcolor{black}{approach to generate worst-case additive noise} that employs norm constraints, in line with the attack strategies utilized in image classification. This approach aims to produce a form of worst-case imperceptible additive noise against a reconstructor in the image domain. Given a perturbation budget $\epsilon>0$, the worst-case additive perturbations can be obtained using the following optimization problem. 
\textcolor{black}{%
\begin{equation}
\label{eqn: attack}
\underset{\|\bm{\delta}\|_\infty \leq \epsilon}{\max}~ \mathcal{L}\Big(\MoDL_\theta (\mathbf{A}^H \mathbf{y}), \MoDL_\theta ( \mathbf{A}^H(\mathbf{y}+\bm{\delta}) ) \Big)\:, 
\end{equation}
}

where $\|.\|_\infty$ is the $\ell_{\infty}$ norm and $\mathcal{L}$ is a differentiable loss function that computes the reconstruction loss. Given the original image $\mathbf{x}^*$, generating the perturbations can also be achieved by replacing the first argument of $\mathcal{L}$ in \eqref{eqn: attack} with $\mathbf{x}^*$. A solution of \eqref{eqn: attack} can be obtained using the Projected Gradient Descent (PGD) method~\cite{madry2017towards}. In this paper, we also use \textcolor{black}{$\mathbf{z}_\textrm{pert} = \mathbf{A}^H(\mathbf{y}+\bm{\delta}) = \mathbf{A}^H\mathbf{y}_{\textrm{pert}}$} which relates perturbations in k-space and image space.

\textcolor{black}{In addition to the worst-case perturbations, random/realistic additive measurement noise could also impact the performance of a reconstructor. }

\subsubsection{{Training/Testing Sampling Protocol \& Undersampling Rate Disparities}}

In addition to additive perturbations, the study presented in \cite{li2023smug} underscores an additional potential source of instability that MoDL (\textcolor{black}{and other DL-based reconstructors}) may face during testing. This source stems from changes in the measurement sampling rate, leading to perturbations in the sparsity of the sampling mask within $\mathbf{A}$ \cite{antun2020instabilities}. Furthermore, in this paper, we consider another variation that these NNs could encounter during the testing phase, involving a shift or variation in the k-space sampling locations within the matrix $\mathbf{M}$, resulting in the construction of a nonidentical forward operator for testing. 
For this case, $\mathbf{z}_\textrm{pert} = \mathbf{A}^H_\textrm{test}\mathbf{y}$, where $\mathbf{A}_\textrm{test} \neq \mathbf{A}$. 

\textcolor{black}{We remark that ensuring the robustness of a reconstruction model to variations in the sampling protocol,  undersampling rate, scan contrast, etc., is crucial as it mitigates the need for re-training to all possible practical scenarios and variations, common in imaging. Re-training models for new setups is expensive. Moreover, the relatively limited training data availability (which requires fully-sampled measurements as labels in supervised learning) in reconstruction applications also warrant learning models that can still be significantly robust.}

\subsubsection{{Unseen Anatomies \& Pathologies at Testing Time}}

\textcolor{black}{A lesion (or anatomy changes) denotes an anomaly,  or impairment within a tissue or organ of the body, arising from diverse factors such as injuries, diseases, or pathological conditions. In the medical domain, the term commonly characterizes regions of abnormal or diseased tissue, observed through MR imaging. In this paper, we study the practical case where the  DL-based image reconstructor is trained on some data points, but tested with measurements with unseen lesions. }

\begin{figure}[t]
    \centering
    \begin{tabular}{ccc}
        \scriptsize{PSNR = 30.8 dB} & 
        \scriptsize{PSNR = 23.21 dB} &
        \scriptsize{PSNR = 22.18 dB} \\
        
        \includegraphics[width=.2\linewidth,angle=180]{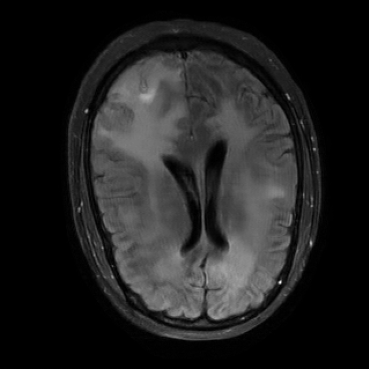} &
        \includegraphics[width=.2\linewidth,angle=180]{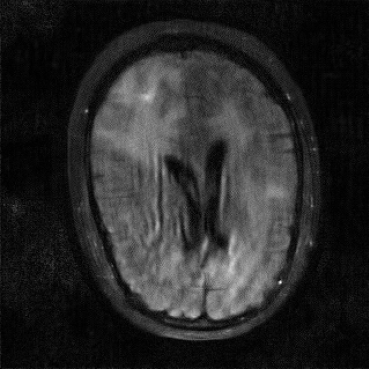} &
        \includegraphics[width=.2\linewidth,angle=180]{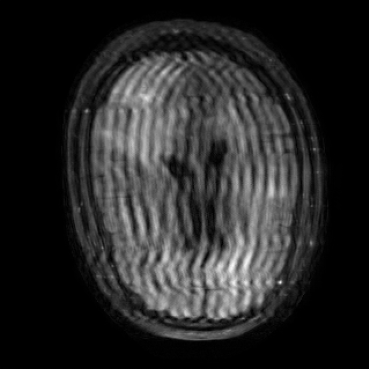} \\
        
        \scriptsize{(a)} & 
        \scriptsize{(b)} &  
        \scriptsize{(c)} \\
    \end{tabular}
    
    \begin{tabular}{cc}
        \scriptsize{PSNR = 24.15 dB} &
        \scriptsize{PSNR = 27.26 dB} \\
        
        \includegraphics[width=.2\linewidth,angle=180]{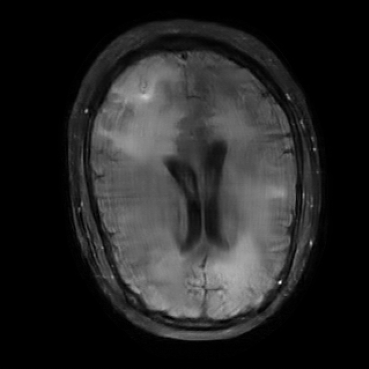} &
        \includegraphics[width=.2\linewidth,angle=180]{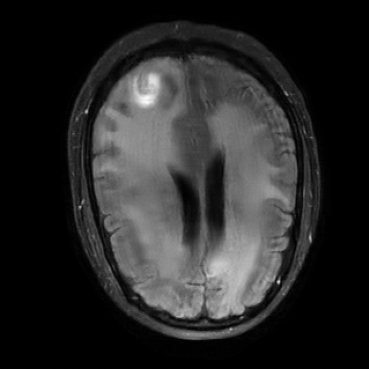} \\
        
        \scriptsize{(d)} &
        \scriptsize{(e)} \\
    \end{tabular}
    
    \caption{\small{\textcolor{black}{Vulnerabilities and generalization challenges of DL-based MRI reconstruction models, demonstrated by evaluating a trained MoDL reconstructor (trained at 4x undersampling) with the considered cases in Section~\ref{sec: Vul}}. (a) Reconstructed image from clean measurements. (b) Reconstructed image from measurements with worst-case additive perturbations \textcolor{black}{(Equation~\eqref{eqn: attack} with $\epsilon = 0.02$)}. (c) Reconstructed image from measurements with 2x undersampling rate during testing. (d) Reconstructed image from \textcolor{black}{a different test time sampling mask with 4x undersampling}. \textcolor{black}{(e) Reconstructed image from measurements with an unseen lesion during testing}.}}
    \label{fig: weaknesses of MoDL}
    \vspace*{-6mm}
\end{figure}

Figure~\ref{fig: weaknesses of MoDL} illustrates reconstructed images from the instabilities \textcolor{black}{and the generalization challenges} considered in this paper.

\subsection{Score-based Diffusion Models}


\textcolor{black}{The Bayesian framework of DMs, introduced in \cite{ho2020denoising,sohl2015deep}, consists of a discrete Markov Chain. The forward direction is constructed by sampling from $p(\mathbf{z}_i \mid \mathbf{z}_{i-1}) = \mathcal{N}(\mathbf{z}_i~;~\sqrt{1-\beta_i}\mathbf{z}_{i-1}, \beta_i\mathbf{I})$, where $\beta_i \in (0,1)$ is an entry of a sequence of monotonically increasing positive noise scales w.r.t. $i$.}  

\textcolor{black}{\textcolor{black}{Score-based DM was introduced in~\cite{song2020score} and was shown to be equivalent to the Bayesian framework}. Score-based DMs can be formulated by the following forward and reverse Stochastic Differential Equations (SDEs). 
\begin{equation}
\label{eqn: SCORE-BASED forward general}
d\mathbf{z} = \textbf{f}(\mathbf{z},t) dt + g(t) d\mathbf{w}\:,
\end{equation}
\begin{equation}
\label{eqn: SCORE-BASED backward general}
d\mathbf{z} = \big[ \textbf{f}(\mathbf{x},t) - g^2(t) \nabla_{\mathbf{z}}\log p_t(\mathbf{z}) \big]  dt + g(t) d\Bar{\mathbf{w}} \:,
\end{equation}
where $\textbf{f}$ and $g$ are the drift and diffusion coefficients, respectively. $t$ spans the interval $[0,1]$ and represents the time index. $d\mathbf{w}$ and $d\Bar{\mathbf{w}}$ represent standard Brownian motion evolving forward and backward in time, respectively. The term $p_t(\mathbf{z})$ denotes the distribution of $\mathbf{z}$ at time $t$, while $\nabla_{\mathbf{z}}\log p_t(\mathbf{z})$ represents the score function. By employing the formulation of the Variance Exploding (VE) SDE (VE-SDE) \cite{song2020score}, for which $\textbf{f} = \textbf{0}$ and $g(t) = \sqrt{d\sigma^2(t)/dt}$, we can re-write the forward and reverse SDEs as 
}

\begin{equation}
\label{eqn: SCORE-BASED forward}
d\mathbf{z} = \sqrt{\frac{d\sigma^2(t)}{dt}}d\mathbf{w} \:,
\end{equation}
\begin{equation}
\label{eqn: SCORE-BASED backward}
d\mathbf{z} = -\frac{d\sigma^2(t)}{dt} \nabla_{\mathbf{z}}\log p_t(\mathbf{z}) dt+ \sqrt{\frac{d\sigma^2(t)}{dt}}d\Bar{\mathbf{w}}\:.
\end{equation}
In Equations \eqref{eqn: SCORE-BASED forward} and \eqref{eqn: SCORE-BASED backward}, function $\sigma(t) = \sigma_l(\sigma_u / \sigma_l)^t$ is a monotonically increasing function w.r.t. $t$, where $\sigma_l\in (0,1)$ and $\sigma_u >1$ are constants. 


The score function is \textcolor{black}{in practice} replaced by a neural network denoted as $s: \mathbb{C}^n \times [0,1] \rightarrow \mathbb{C}^n$, parameterized by $\phi$, which is trained using the denoising score matching technique \cite{chung2022score} as
\begin{equation}
\label{eqn: DM training}
\underset{\phi}{\min}~ \mathbb{E} \Bigg[ \left\| \sigma(t)s_\phi(\mathbf{z}(t),t) -  \frac{\mathbf{z}(t) - \mathbf{z}}{\sigma(t)}   \right\|^2 \Bigg]   \:.
\end{equation}
The expectation in \eqref{eqn: DM training} is taken over $t\sim U[0,1]$, $\mathbf{z} \sim p(\mathbf{z})$, and $\mathbf{z}(t) \sim \mathcal{N}(\mathbf{z},\sigma(t)\mathbf{I})$, where $p(\mathbf{z})=p_0(\mathbf{z})$ is the distribution of the training data.\par 

\textcolor{black}{Having obtained a trained DM with parameters $\phi$}, the task of sampling $\hat{\mathbf{z}}(0)$ at the time instant $t=0$ is realized through the solution of the reverse process SDE in \eqref{eqn: SCORE-BASED backward}. In this step, the score function is substituted with the learned function $s_\phi$. 
There exist various techniques for sampling from DMs, which involve solving the reverse SDE in \eqref{eqn: SCORE-BASED backward}. In this paper, the Euler method \cite{platen2010numerical} and the Predictor-Corrector (PC) scheme \cite{allgower2012numerical} are used. Following the work in \cite{chung2022score}, a data consistency step is considered to allow sampling from the conditional distribution $p(\mathbf{z} | \mathbf{y})$, 
In practice, the continuous time index $t\in [0,1]$ is discretized into $i\in [N_{{r}}]$, where $[N_{{r}}]:=\{1,\dots,N_{r}\}$. The PC sampling technique consists of $N_r$ prediction reverse steps. In each prediction iteration, $M_r$ correction steps are required \cite{song2020score}. The full procedure is outlined in Algorithm~\ref{alg: PC with DC}. 

\textcolor{black}{We remark that while for the major part of the paper, we use Algorithm~\ref{alg: PC with DC}, our proposed method (which we will discuss in the next section) can be adapted to accelerated DM-based samplers. An example of such samplers is the Denoising Diffusion Implicit Models (DDIM) \cite{songdenoising} as will be illustrated in Section~\ref{sec: exp}.}

\begin{algorithm}[t]
\small
\caption{Predictor-Corrector Sampling with DC \cite{chung2022score}}
\textbf{Input}: Image $\mathbf{z}= \mathbf{A}^H\mathbf{y}$, trained DM $s_\phi$, discretized time step $N_{r}$, and noise schedule $\epsilon_i$.  \\
\textbf{Function}: $\hat{\mathbf{z}} = \SDE\big(s_\phi(\mathbf{z}(N_{r}),N_{r}), \mathbf{y}, \mathbf{A} ,N_{r},0\big)$. \\
\vspace{1mm}
\small{ 1:}  \textbf{Initialize} $\mathbf{z}(N_{r})\sim \mathcal{N}(0,\sigma^2(N_r) \mathbf{I})$. \\
\vspace{1mm}
\small{ 2:}  \textbf{For} $i\in \{N_r-1,\dots,0\}$ \textbackslash\textbackslash Prediction  \\
\vspace{1mm}
\small{ 3:} \hspace{2mm}  $\mathbf{z}'(i) \leftarrow \mathbf{z}(i+1) + (\sigma^2(i+1)-\sigma^2(i))s_\phi(\mathbf{z}(i+1),i+1)$ \\
\vspace{1mm}
\small{ 4:} \hspace{2mm}  $\mathbf{z}(i) \leftarrow \mathbf{z}'(i) + \sqrt{\sigma^2(i+1)-\sigma^2(i)} \eta,~\eta\sim \mathcal{N}(0,\mathbf{I})$ \\
\vspace{1mm}
\small{ 5:} \hspace{2mm}  $\mathbf{z}(i) \leftarrow \mathbf{z}(i) +  \mathbf{A}^H (\mathbf{y} - \mathbf{A}\mathbf{z}(i)) $ \textbackslash\textbackslash Data Consistency\\
\vspace{1mm}
\small{ 6:}  \hspace{2mm} \textbf{For} $M_r$ steps \textbf{do} \textbackslash\textbackslash Correction\\
\vspace{0.5mm}
\small{ 7:} \hspace{4mm}  $\mathbf{z}'(i) \leftarrow \mathbf{z}(i) + \epsilon_i s_\phi(\mathbf{z}(i),i)$ \\
\vspace{1mm}
\small{ 8:} \hspace{4mm}  $\mathbf{z}'(i) \leftarrow \mathbf{z}'(i) + \sqrt{2\epsilon_i}~\eta,~\eta\sim \mathcal{N}(0,\mathbf{I})$ \\
\vspace{1mm}
\small{ 9:} \hspace{4mm}  $\mathbf{z}(i) \leftarrow \mathbf{z}'(i) + \mathbf{A}^H (\mathbf{y} - \mathbf{A}\mathbf{z}(i)) $ \textbackslash\textbackslash Data Consistency\\
\vspace{1mm}
\small{10:} \hspace{0mm}  $\hat{\mathbf{z}} =\mathbf{z}(0)$ \\

\vspace{-4.0mm}
\label{alg: PC with DC}
\end{algorithm}


\section{Diffusion Purification for Robust DL-based MRI Reconstruction}
\label{sec: prop method}
In this section, we begin by outlining the key components of the proposed Diffusion Purification (DP) pipeline. Subsequently, we introduce our approach for obtaining the PST step. Following that, we elaborate on our fine-tuning strategy for MoDL, leveraging the purified samples. \textcolor{black}{We dub our method as \textbf{RO}bust \textbf{D}L-based MR\textbf{I} with Diffusion Purificati\textbf{O}n (\textbf{RODIO}).} 



\subsection{DM-based Purification}

Here, we present our DP approach, which consists of the following two stages.

\noindent \textbf{Diffusion Stage:} Given measurements $\mathbf{y}$, let $\mathbf{z}_\textrm{pert}$ denote the perturbed version of $\mathbf{z}=\mathbf{A}^H\mathbf{y}$. As illustrated in the previous section, this perturbed version can be due to various reasons such as 
random measurement noise, not well-modeled noise and artifacts (e.g., it may make sense to consider worst-case additive noise (from \eqref{eqn: attack})), and different k-space undersampling factors or sampling patterns/masks at testing time.

The first stage of the DP approach involves diffusing $\mathbf{z}(0) = \mathbf{z}_\textrm{pert}$ from $t=0$ to $t=t^*$, where $t^*\in (0,1)$ indicates the diffusion time index at which the forward process stops. We term $t^*$ as the Process-Switching Time (PST) step. The PST step and $\sigma(\cdot)$ control the amount of noise added to $\mathbf{z}_\textrm{pert}$. This stage corresponds to 
\begin{equation}
\label{eqn: method step 1}
\mathbf{z}_\textrm{pert}(t^*) = \mathbf{z}_\textrm{pert} + \sqrt{\sigma^2(t^*) - \sigma^2(0)} \eta_{t^*}\:,  \eta_{t^*} \sim \mathcal{N}(0,\mathbf{I}) \:.
\end{equation}
\noindent \textbf{Purification Stage:} After obtaining the diffused perturbed image, denoted as $\mathbf{z}_\textrm{pert}(t^*)$, the objective of the second step is to derive the purified sample, denoted as $\mathbf{z}^{\textrm{pur}}_\textrm{pert}$, from  $\mathbf{z}_\textrm{pert}(t^*)$. This is achieved by employing the PC reverse process with data consistency (DC). In other words, we use the PC with DC procedure in Algorithm~\ref{alg: PC with DC} as:
\begin{equation}
\label{eqn: method step 2}
\mathbf{z}^{\textrm{pur}}_\textrm{pert}(0) = \SDE\big(s_\phi(\mathbf{z}_\textrm{pert}(t^*),t^*), \mathbf{y}_\textrm{pert},\mathbf{A},t^*,0\big)\:.
\end{equation}
In practice, we use $N_{t^*}$, which represents the discrete PST step. We remark that $N_{t^*}$ is less than the total number of available steps in standard sampling reverse process $N_r$. Algorithm~\ref{alg: DP algorithm} illustrates the diffusion purification procedure.\par
\hspace{0.33cm}
\begin{figure*}  
\centering
\includegraphics[width=15cm]{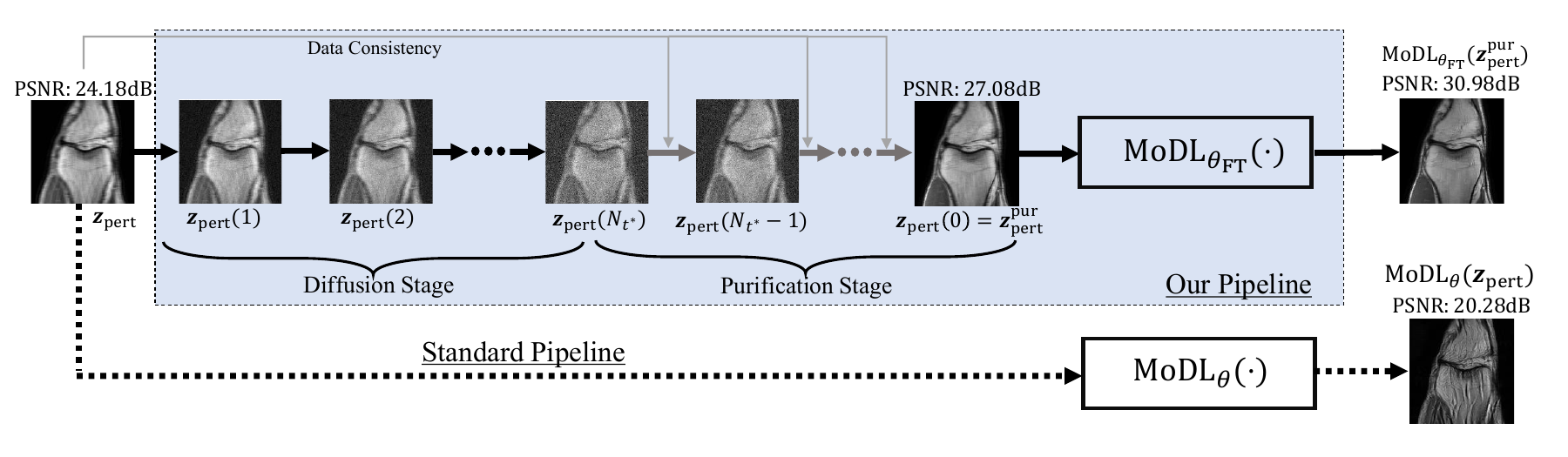}
\vspace{-0.5cm}
\caption{\small{A schematic block diagram illustrating the standard pipeline and our proposed reconstruction pipeline. The functions $\MoDL_{\theta}(\cdot)$ and $\MoDL_{\theta_{\textrm{FT}}}(\cdot)$ represent the application of the standard pre-trained MoDL procedure and our `pre-trained+fine-tuned' robust MoDL procedure, respectively. \textcolor{black}{Here, MoDL can be replaced with other DL-based reconstruction models.}}}
\label{fig: BD}
\vspace{-0.4cm}
\end{figure*}

\noindent \textbf{Intuition:} Starting with a perturbed image $\mathbf{z}_\textrm{pert}$, which is assumed to be drawn from distribution $q(\mathbf{z})$, our approach initiates with $\mathbf{z}(0) = \mathbf{z}_{\textrm{pert}}$ and gradually introduces noise. If the aliased image $\mathbf{z}$ follows a distribution $p(\mathbf{z})$, then as $t\rightarrow 1$, these two distributions will get closer. This signifies that the perturbations are progressively diminishing due to the incremental noise incorporated during the forward process of \eqref{eqn: method step 1}. \textcolor{black}{To emphasize this point, we present the following Theorem, whose proof is deferred to the Appendix.} \par
\hspace{0.33cm}


\begin{mytheorem}[]
\label{th: KL}
Let $p_{t}(\mathbf{z})$ and $p_{0t}(\mathbf{z}(t)\mid \mathbf{z})$ be the distribution and the conditional distribution of $\mathbf{z}(t)$ given the VE-SDE forward process of \eqref{eqn: SCORE-BASED forward} starts at the unperturbed image $\mathbf{z}$. Similarly, let $q_{t}(\mathbf{z})$ and $q_{0t}(\mathbf{z}(t)\mid \mathbf{z}_{\textrm{pert}})$ be the distribution and the conditional distribution of $\mathbf{z}(t)$ given the VE-SDE forward process of \eqref{eqn: SCORE-BASED forward} starts at the perturbed image $\mathbf{z}_{\textrm{pert}} = \mathbf{A}^H \mathbf{y}_{\textrm{pert}}= 
\mathbf{A}^H (\mathbf{y}+\bm{\delta})$. Then, as $t$ moves forward from $t=0$ to $t=1$: 

\begin{enumerate}[left=0pt]
    \item The KL divergence between $p_{0t}$ and $q_{0t}$, defined in \eqref{eqn: KL theorem}, monotonically decreases.
\begin{equation}
\label{eqn: KL theorem}
D_{\textrm{KL}}(p_{0t}\mid\mid q_{0t}) = \frac{\|\mathbf{A}^H \bm{\delta}\|^2}{2(\sigma^2(t)-\sigma^2(0))}\:, t\in (0,1] \:.
\end{equation}

\item The KL divergence between $p_{t}$ and $q_{t}$ monotonically decreases, i.e., 
\begin{equation}
\label{eqn: KL theorem 2}
\frac{dD_{\textrm{KL}}(p_{t}\mid\mid q_{t})}{dt} \leq 0\:. 
\end{equation}
\end{enumerate}
\end{mytheorem}

\textcolor{black}{It is important to highlight that our Theorem uses the VE-SDE, where the probability distributions are from the standard Bayesian framework of DMs \cite{ho2020denoising}}.

\begin{algorithm}[t]
\small
\caption{Diffusion Purification}
\textbf{Input}: Perturbed measurements $\mathbf{y}_\textrm{pert}$, operator $\mathbf{A}$, trained DM $s_\phi$, and PST step $N_{t^*}$  \\
\textbf{Function}: $\mathbf{z}^{\textrm{pur}}_\textrm{pert} = \DP_{\phi}(\mathbf{y}_\textrm{pert},\mathbf{A}, N_{t^*})$. \\
\vspace{1mm}
\small{ 1:}  \textbf{Initialize} $\mathbf{z}(0) = \mathbf{z}_\textrm{pert}$\\
\vspace{1mm}
\small{ 2:}  \textbf{For} $i\in \{1,\dots,N_{t^*}\}$ \textbackslash\textbackslash Diffusion steps\\
\vspace{1mm}
\small{ 3:} \hspace{2mm}  \textbf{Obtain} $\mathbf{z}(i) \leftarrow \mathbf{z}(i-1) + \sqrt{\sigma^2(i) - \sigma^2(i-1)}~ \eta, \eta\sim\mathcal{N}(0,\mathbf{I})$ \\
\vspace{1mm}
\small{ 4:}  \textbf{For} $i\in \{N_{t^*},\dots,1\}$ \textbackslash\textbackslash Purification steps\\
\vspace{1mm}
\small{ 5:} \hspace{2mm}  \textbf{Obtain} $\mathbf{z}(i-1) \leftarrow \SDE(s_\phi(\mathbf{z}(i),i),\mathbf{y}_\textrm{pert},\mathbf{A}, i, i-1)$ \\
\vspace{1mm}
\small{ 6:}  \textbf{Obtain} $\mathbf{z}^{\textrm{pur}}_\textrm{pert} = \mathbf{z}(0)$. \\
\vspace{1mm}
\vspace{-4.2mm}
\label{alg: DP algorithm}
\end{algorithm}
%






\subsection{Selection of the Process-Switching Time Step}

Here, we present an \textit{approximate} method to obtain $t^*<1$ (or $N_{t^*}<N_r$) based on the Maximum Mean Discrepancy (MMD) metric \cite{gretton2006kernel}. \textcolor{black}{The MMD metric measures the dissimilarity between two distributions by comparing their mean embedding in a reproducing kernel Hilbert space. It is commonly employed in machine learning and statistics for various tasks, including domain adaptation \cite{guan2021domain} and kernel methods \cite{hofmann2008kernel}.}\par

We utilize the MMD metric to approximately quantify the empirical distribution shift between the original distribution $p(\mathbf{z})$ and the perturbed images' distribution $q(\mathbf{z})$. During the forward diffusion process, let $Z(i)$ and $Z_\textrm{p}(i)$ (with $|Z(i)|=|Z_{\textrm{p}}(i)|$) represent the set of unperturbed and perturbed images, respectively, at discrete time step $i$, where $|\cdot|$ denotes the cardinality of a set. Since we lack access to the exact distributions, we can approximate $\mmd(p_i,q_i)$ using empirical distributions as follows:
\begin{equation}
\begin{aligned} \label{eqn: MMD}
\mmd(p_i,q_i) \approx C \Big( \sum_{\mathbf{z}(i), \mathbf{z}'(i)\in Z(i), \mathbf{z}(i)\neq \mathbf{z}'(i)} k(\mathbf{z}(i),\mathbf{z}'(i)) 
+\\ \sum_{\mathbf{z}(i),\mathbf{z}'(i)\in Z_{\textrm{p}(i)}, \mathbf{z}(i)\neq \mathbf{z}'(i)} k(\mathbf{z}(i),\mathbf{z}'(i)) \Big) \\
\hspace{-2cm}- \frac{2}{|Z(i)|^2} \sum_{\mathbf{z}(i)\in Z(i),\mathbf{z}'(i)\in Z_{\textrm{p}}(i)} k(\mathbf{z}(i),\mathbf{z}'(i))
\:,   
\end{aligned}    
\end{equation}
%
where $C = 1/(|Z(i)|(|Z(i)|-1))$ is used for brevity, and $k(\mathbf{z}(i),\mathbf{z}'(i)) = \exp(-\|\mathbf{z}(i)-\mathbf{z}'(i)\|^2/2v^2)$ 
%
is the Gaussian kernel parameterized by $v>0$.\par



Considering the balance between purifying additive perturbations (achieved with a larger $t^*$) and preserving global structures (achieved with a smaller $t^*$) within perturbed samples, there exists an ideal value of $t^*$ that yields a robust reconstruction accuracy. In the case of the worst-case additive perturbations, the changes are usually small and can be rectified with a small $t^*$. It was shown in \cite{pmlr-v162-nie22a} that the most efficient choice of $t^*$ related to adversarial robustness tends to be on the smaller side. As such, our objective is to find the minimum value of $i\in [N_r]$ for which $\mmd(p_i,q_i)\approx 0$. Consequently, we formulate the following optimization problem to determine the near-optimal discrete PST step, $N_{t^*}$.
\begin{equation}
\begin{aligned} \label{eqn: PST}
N_{t^*} := \big\{ \underset{i\in [N_r]}{\arg \min}~ i ~~  \text{s.t.} ~~ \mmd(p_i,q_i)=0 \big\}\:.
\end{aligned}    
\end{equation}
In order to obtain the solution of \eqref{eqn: PST}, it is required to perform the forward diffusion (steps 2 and 3 in Algorithm~\ref{alg: DP algorithm}) on the unperturbed and perturbed samples until the constraint is satisfied.\par

Since we have knowledge of the source of perturbations that allows us to obtain $Z_{\textrm{p}}$ from $Z$, we remark that the PST step selection method we propose can be applied to any diffusion purification task.



\subsection{Fine-tuning with Purified Perturbed Examples}

In this subsection, drawing inspiration from the widely used `pre-training + fine-tuning' approach \cite{zoph2020rethinking,salman2020denoised}, we propose fine-tuning the parameters of MoDL, which are obtained through the process outlined in Section~\ref{sec: prel}.A, using contaminated purified examples.

We start with pre-trained parameters $\theta$, and utilize noised purified examples for fine-tuning. Let $\theta_{\textrm{FT}}$ represent the fine-tuned parameters specific to MoDL. Initially, we set $\theta_{\textrm{FT}}$ equal to $\theta$. Then, for each measurement $\mathbf{y}$ within dataset $D$, we generate a \textcolor{black}{noisy version} of the aliased reconstruction, $\mathbf{A}^H(\mathbf{y}+\mathbf{v})$, where $\mathbf{v}$ is drawn from a normal distribution $\mathcal{N}(0,\sigma_{\textrm{FT}}\mathbf{I})$. Subsequently, for every $(\mathbf{y},\mathbf{x})$, we follow the procedure outlined in \cite{aggarwal2018modl}, while initializing $\mathbf{x}_0$ as
\begin{equation}
\label{eqn: fine-tuning MoDL init}
\mathbf{x}_0 = \DP_{\phi}(\mathbf{y}+\mathbf{v},\mathbf{A},N_{t^*}) \:.
\end{equation}
Having trained $\theta_{\textrm{FT}}$ that maps $\mathbf{x}_0$ to fully-sampled reconstructions, at the testing phase, the robust MoDL MRI reconstruction using diffusion purification is represented in Algorithm~\ref{alg: Robust MoDL Pipeline}. A block diagram of the proposed approach is given in Figure~\ref{fig: BD}.

\begin{algorithm}[t]
\small
\caption{\textcolor{black}{Our Robust MRI Reconstruction Pipeline (RODIO) with MoDL}}
\textbf{Input}: Perturbed measurements $\mathbf{y}_\textrm{pert}$, operator $\mathbf{A}$, trained DM $s_\phi$, PST step $N_{t^*}$, number of unrolling steps $N$, and fine-tuned MoDL parameters $\theta_{\textrm{FT}}$.  \\
\textbf{Output}: Reconstructed image after purification $\mathbf{x}$. \\
\vspace{1mm}
\small{ 1:} \textbf{Obtain} $\mathbf{z}^{\textrm{pur}}_\textrm{pert} = \DP_{\phi}(\mathbf{y}_\textrm{pert}, \mathbf{A} , N_{t^*})$. \\
\vspace{1mm}
\small{ 2:}  \textbf{Initialize} MoDL reconstructed image as $\mathbf{x}_0 = \mathbf{z}^{\textrm{pur}}_\textrm{pert}$ \\
\vspace{1mm}
\small{ 3:}  \textbf{For} $j\in \{0,\dots,N-1\}$ \textbackslash\textbackslash MoDL unrolling steps \\
\vspace{1mm}
\small{ 4:} \hspace{2mm}  \textbf{Obtain} $\mathbf{z}_j \leftarrow f_{\theta_{\textrm{FT}}} (\mathbf{x}_j)$  \\ 
\vspace{1mm}
\small{ 5:} \hspace{1.92mm}  \textbf{Obtain} $\mathbf{x}_{j+1} \leftarrow (\mathbf{A}^H \mathbf{A} +\lambda \mathbf{I})^{-1}(\mathbf{z}^{\textrm{pur}}_\textrm{pert}+\lambda \mathbf{z}_j)$  \\
\vspace{1mm}
\small{ 6:}  \textbf{Obtain} $\mathbf{x} \leftarrow \mathbf{x}_{N}$  \\
\vspace{-3.0mm}
\label{alg: Robust MoDL Pipeline}
\end{algorithm}


\textcolor{black}{We emphasize that while our primary focus is on the formulation of MoDL under which we develop our proposed approach, in the last subsection of our experimental results, we demonstrate the versatility of our approach by showcasing its applicability to other DL-based supervised MRI reconstruction models.}






\section{Experimental Results}
\label{sec: exp}

In this section, we start by \textcolor{black}{illustrating} our experimental setup, \textcolor{black}{baselines}, and the \textcolor{black}{instability} sources \textcolor{black}{and the generalization challenges} considered in this work. Subsequently, we present results for the process-switching time (PST) step selection through our MMD-based method. Following this, we present the primary results showcasing the robustness of our approach. \textcolor{black}{Furthermore, we present visualizations illustrating knee and brain MRI reconstructions.} 



\subsection{Experimental Setup}

In the case of MoDL, we employ a configuration with $N = 6$ unrolling steps and a regularization parameter $\lambda = 1$. The architecture of $f_\theta$ is selected as the Deep Iterative Down-Up Network \cite{yu2019deep}. Additionally, we set the convergence threshold for the conjugate gradient optimization used in the data consistency step of \eqref{eqn: MoDL training 2} to $10^{-6}$. In the DM setting, $t\in [0,1]$ is discretized into $500$ steps. We adopt a pre-trained DM model from \cite{chung2022score}, where $\sigma(i)$ is a geometric series selected as $\sigma(i) = 0.01(37800)^{\frac{i}{N_r-1}}$. 
We note that the DM model was trained on the knee training dataset. 
We conduct our experiments on the fastMRI dataset \cite{zbontar2018fastmri}, using 3000 purified images for fine-tuning the pre-trained MoDL network. Additionally, 20 images are reserved for validation, and 64 images are used for testing. Moreover, we use $\sigma_{\textrm{FT}} = 0.01$. The multi-coil image data is acquired using $15$ coils and is cropped to a resolution of $320 \times 320$ pixels for MRI reconstruction. To simulate undersampling of the MRI k-space, we adopt a Cartesian mask with 4x acceleration (equivalent to a 25\% sampling rate). Sensitivity maps for the coils, which are incorporated into the operator $\mathbf{A}$ for all scenarios, are obtained using the BART toolbox \cite{tamir2016generalized}. Rather than employing the root-sum-of-squares reconstruction method, we apply sensitivity map-based reconstruction. The quality of the reconstructed images is evaluated using the Peak Signal-to-Noise Ratio (PSNR) in dB, and the Structural Similarity Index Measure (SSIM), which returns values in $[0,1]$ with $1$ indicating identical images. All the experiments are conducted on a single RTX5000 GPU machine, and our code is made available online\footnote{\tiny{\url{https://anonymous.4open.science/r/adversarial-purification-for-MRI-125B/README.md}}}.\par

%


\noindent{\textbf{Baselines:}} Here, we list the baselines used in our experiments.

\subsubsection{\textcolor{black}{Vanilla DL-based MRI Reconstructors}} \textcolor{black}{Here, we consider standalone MoDL and Recurrent VarNet. These are also incorporated within our proposed framework.}.

\subsubsection{\textcolor{black}{Adversarial Training}} In AT, we implemented a 30-step PGD procedure within its minimax formulation.

\subsubsection{\textcolor{black}{E2E Randomized Smoothing}} For E2E-RS, we introduced Gaussian noise with a standard deviation of $0.01$, and to perform the smoothing operation, we employed 10 Monte Carlo samplings. 


\subsubsection{\textcolor{black}{Score-MRI}} We compare our proposed approach with a diffusion-based method, namely the score-MRI work in \cite{chung2022diffusion}.

\subsubsection{\textcolor{black}{DDS}} \textcolor{black}{We also compare with a more recent diffusion-based method, namely the DDS work in \cite{chung2024decomposed}.} 


\subsubsection{\textcolor{black}{Standalone Diffusion Purification}} \textcolor{black}{We report results of using only the diffusion purifier with data-consistency (Algorithm~2). We use `DP' to refer to this case.}

\subsubsection{\textcolor{black}{LORAKI}} \textcolor{black}{LORAKI is an unsupervised recurrent neural network tailored for MRI reconstruction in k-space, representing a scan-specific method. \textcolor{black}{Here,} we utilize the modification in \cite{Raki}, where a publicly available code is used\footnote{\tiny{\url{https://github.com/geopi1/DeepMRI}}}.} \textcolor{black}{For generating worst-case noise, we use the same approach as in \eqref{eqn: PCDC attack}.}



%
\begin{figure}[t]
\centering
\includegraphics[width=8cm]{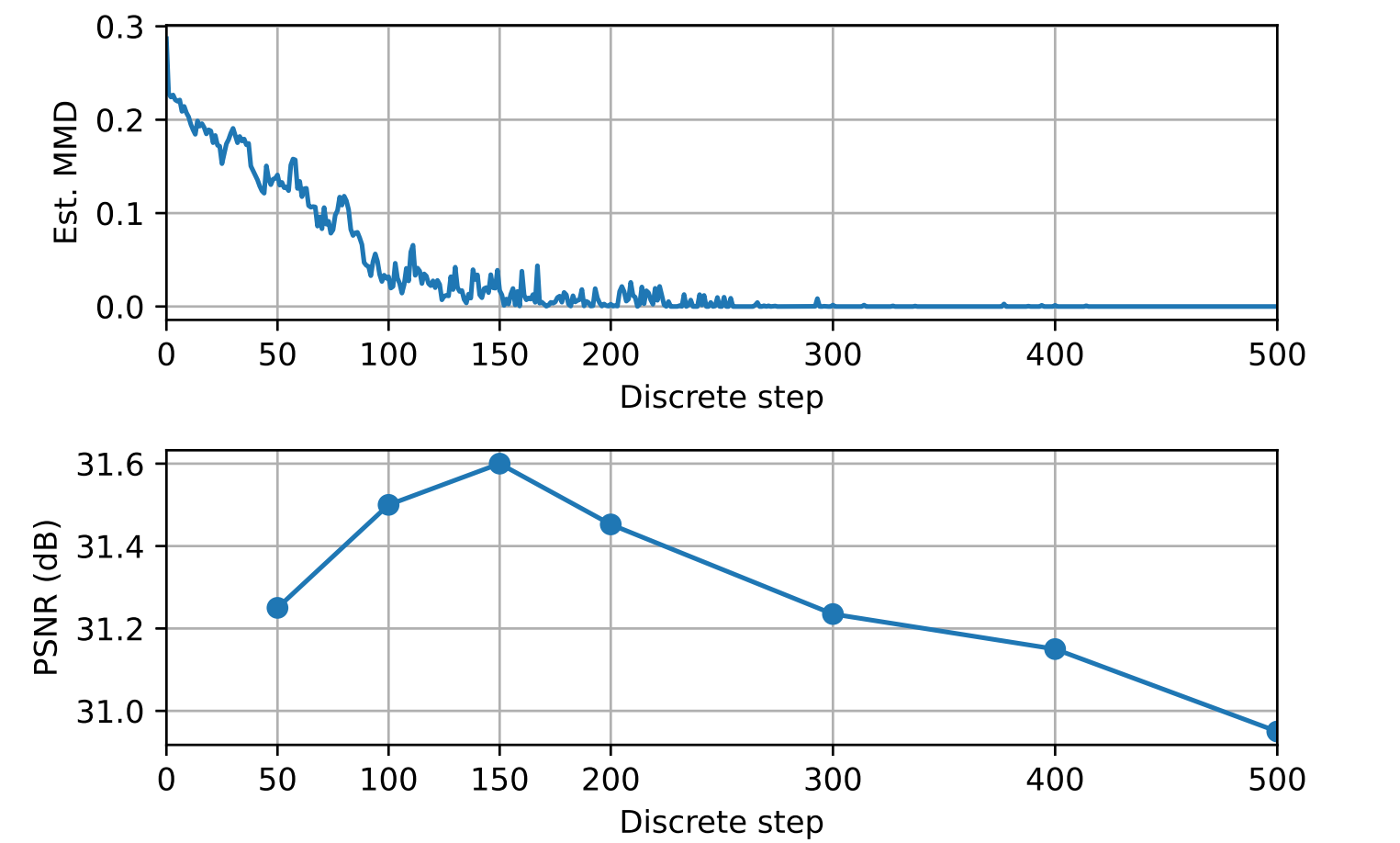}
\vspace{-0.23cm}
\caption{\small{Selection of the PST step using fastMRI dataset (i.e., the main dataset used in major parts of the paper). Estimated MMD (using \eqref{eqn: MMD}) w.r.t. the discrete steps $i\in [N_r]$ (\textit{top}). Ablation study by comparing with the ground truth (\textit{bottom}). }}
\label{fig: tstar ablation}
\vspace{-0.3cm}
\end{figure}
\begin{figure}[t]
\centering
\includegraphics[width=8cm]{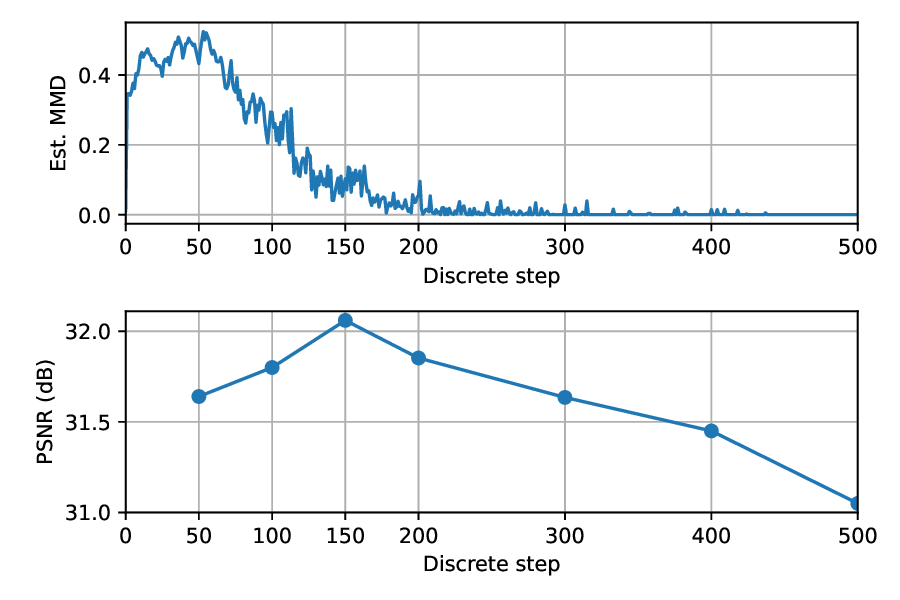}
\vspace{-0.23cm}
\caption{\small{\textcolor{black}{Selection of the PST step for the Stanford 2D FSE Dataset \cite{FSE}. Estimated MMD (using \eqref{eqn: MMD}) w.r.t. the discrete steps $i\in [N_r]$ (\textit{top}). Ablation study by comparing with the ground truth (\textit{bottom}).}}}
\label{fig: tstar ablation 2}
\vspace{-0.3cm}
\end{figure}
\begin{figure}[t]
\centering
\includegraphics[width=7.4cm]{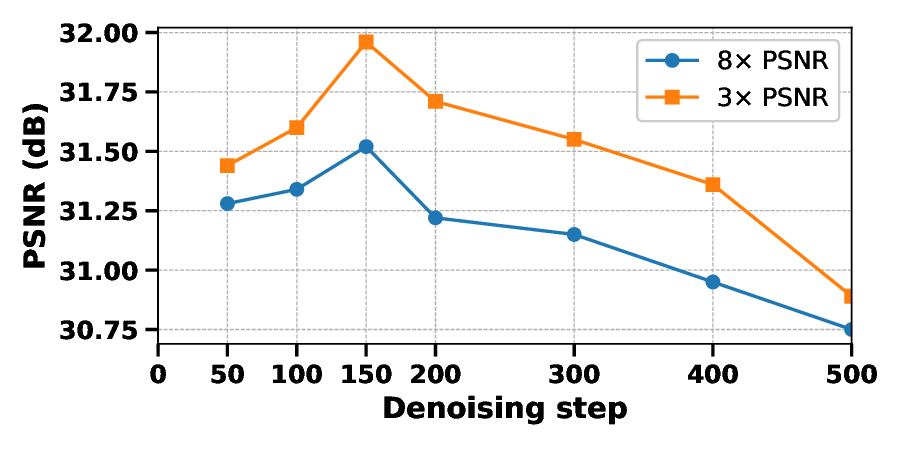}
\vspace{-0.39cm}
\caption{\small{\textcolor{black}{Performance of our method using undersampling factors of 8x and 3x (which are settings different from the training setting, 4x) across different selections of the process switching time (PST).}}}
\label{fig: tstar ablation 3}
\vspace{-0.6cm}
\end{figure}
%
\subsection{Implementation Details for the Sources of Instabilities \& Generalization Settings}




\subsubsection{k-space Additive Noise} \textcolor{black}{Here, we consider} additive perturbations applied to the \textcolor{black}{measurements} $\mathbf{y}$. Recall that for example in the unrolled MoDL, this is both an input and is used in the conjugate gradients (CG) scheme in the data consistency step. We \textcolor{black}{consider} two types of additive noise: a zero-mean \textcolor{black}{complex} Gaussian random vector with a variance of $0.01$, and worst-case additive perturbations. For the latter, we employed two gradient-based \textcolor{black}{optimization} techniques. The first method is the conventional $\ell_\infty$-norm PGD~\cite{madry2017towards} with 30 iterations and a perturbation budget of $\epsilon = 0.004$. The second approach utilizes the advanced momentum-based AUTO attack~\cite{croce2020reliable}, configured similarly to PGD. To generate perturbations using PGD or AUTO, it is necessary to calculate the gradients w.r.t. the input of our model. 


In this paper, we consider an additional case where we apply the method from \cite{pmlr-v162-nie22a} and calculate the gradients to propagate through both MoDL and the SDE of the DP. This represents the worst-case additive perturbations w.r.t. the DP and MoDL. In this case, the perturbations are generated as: 
\begin{equation}
\begin{aligned} \label{eqn: attack e2e}
\underset{\|\bm{\delta}\|_\infty \leq \epsilon}{\max} \mathcal{L}\Big(\MoDL_{\theta_{\textrm{FT}}}(\DP_\phi(\mathbf{A}^H\mathbf{y},N_{t^*})),\\ \MoDL_{\theta_{\textrm{FT}}}(\DP_\phi(\mathbf{A}^H (\mathbf{y}+\bm{\delta}),N_{t^*})) \Big)\:. 
\end{aligned}
\end{equation}

Worst-case additive noise for AT, E2E-RS, Recurrent Var Net and LORAKI are generated using the optimization problem in \eqref{eqn: attack}, with changing the structure of the network. For score-MRI and standalone diffusion purification, we use Equations \eqref{eqn: PCDC attack} and \eqref{eqn: DP attack}, respectively, which are modified versions of \eqref{eqn: attack}.
\begin{equation}
\begin{aligned}
    \label{eqn: PCDC attack}
\underset{\|\bm{\delta}\|_\infty \leq \epsilon}{\max}~ \mathcal{L}\Big(\SDE\big(s_\phi(\mathbf{z}(N_{r}),N_{r}), \mathbf{y}+\boldsymbol{\delta}, \mathbf{A} ,N_{r},0\big), \\
\SDE\big(s_\phi(\mathbf{z}(N_{r}),N_{r}), \mathbf{y}, \mathbf{A} ,N_{r},0\big) \Big)\:. 
\end{aligned}
\end{equation}
\begin{equation}
\begin{aligned}
    \label{eqn: DP attack}
\max_{\|\bm{\delta}\|_\infty \leq \epsilon} \mathcal{L}\big(\DP_\phi(\mathbf{A}^H \mathbf{y},N_{t^*}), \DP_\phi(\mathbf{A}^H (\mathbf{y}+\bm{\delta}),N_{t^*}) \big)\:.  
\end{aligned}
\end{equation}
\begin{figure*}
    \centering
    \includegraphics[width=0.81\textwidth]{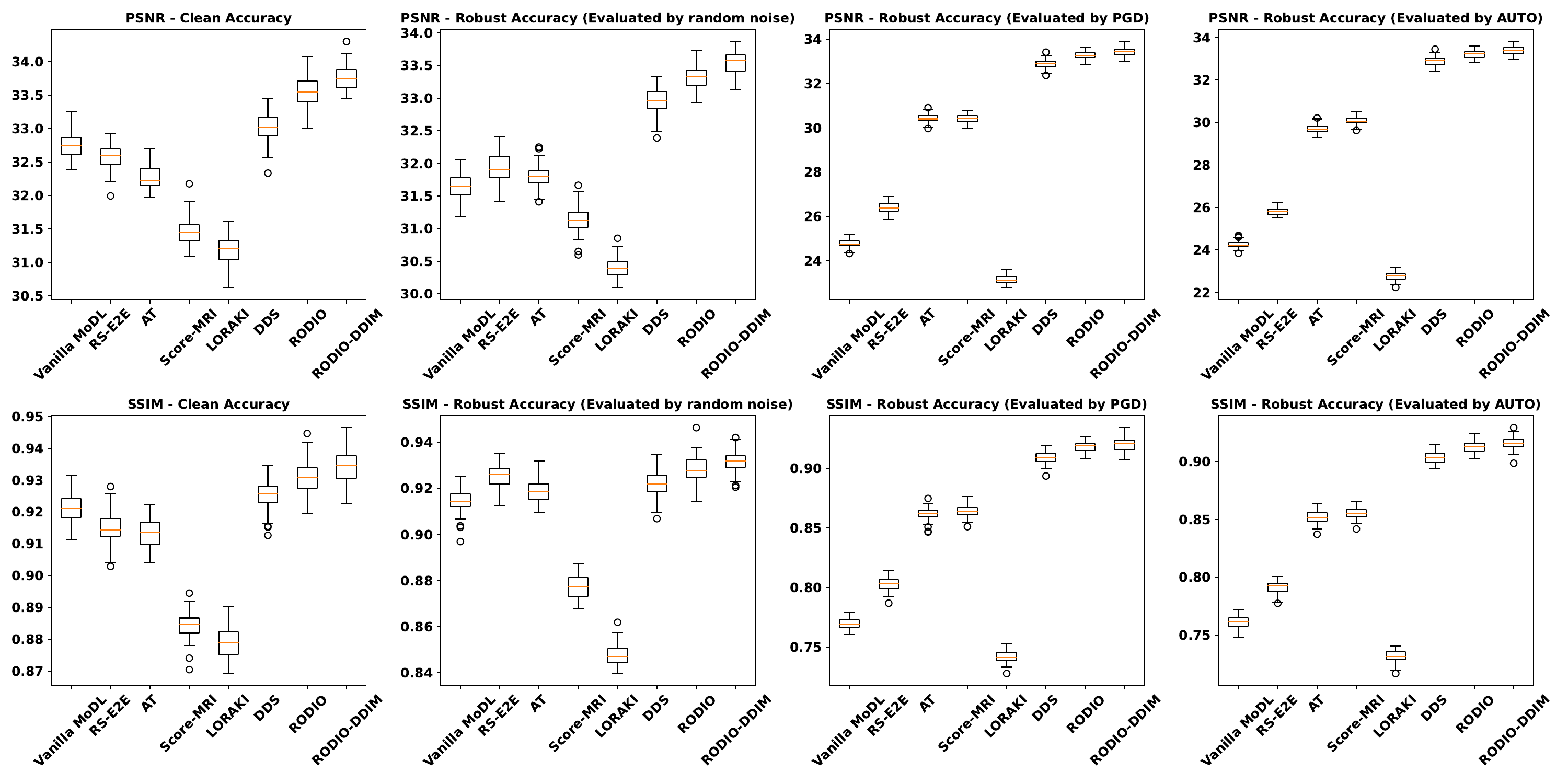}
    \vspace{-0.35cm}
    \caption{\small{ \textcolor{black}{Reconstruction accuracy box plots for the \textbf{knee} fastMRI dataset with 4x Acceleration factor. The additive Gaussian random noise of the second column plots is obtained using variance of $0.01$. The worst-case additive noise of the third and fourth columns are obtained using PGD and AUTO methods with $\epsilon = 0.02$. \textcolor{black}{Here, RODIO (resp. RODIO-DDIM) corresponds to using the sampler in Algorithm~\ref{alg: PC with DC} (resp. DDIM with measurement consistency) for diffusion purification.} }}}
    \label{fig: box plot knee}
\end{figure*}
\begin{figure}[t]
\centering
\includegraphics[width=7.5cm]{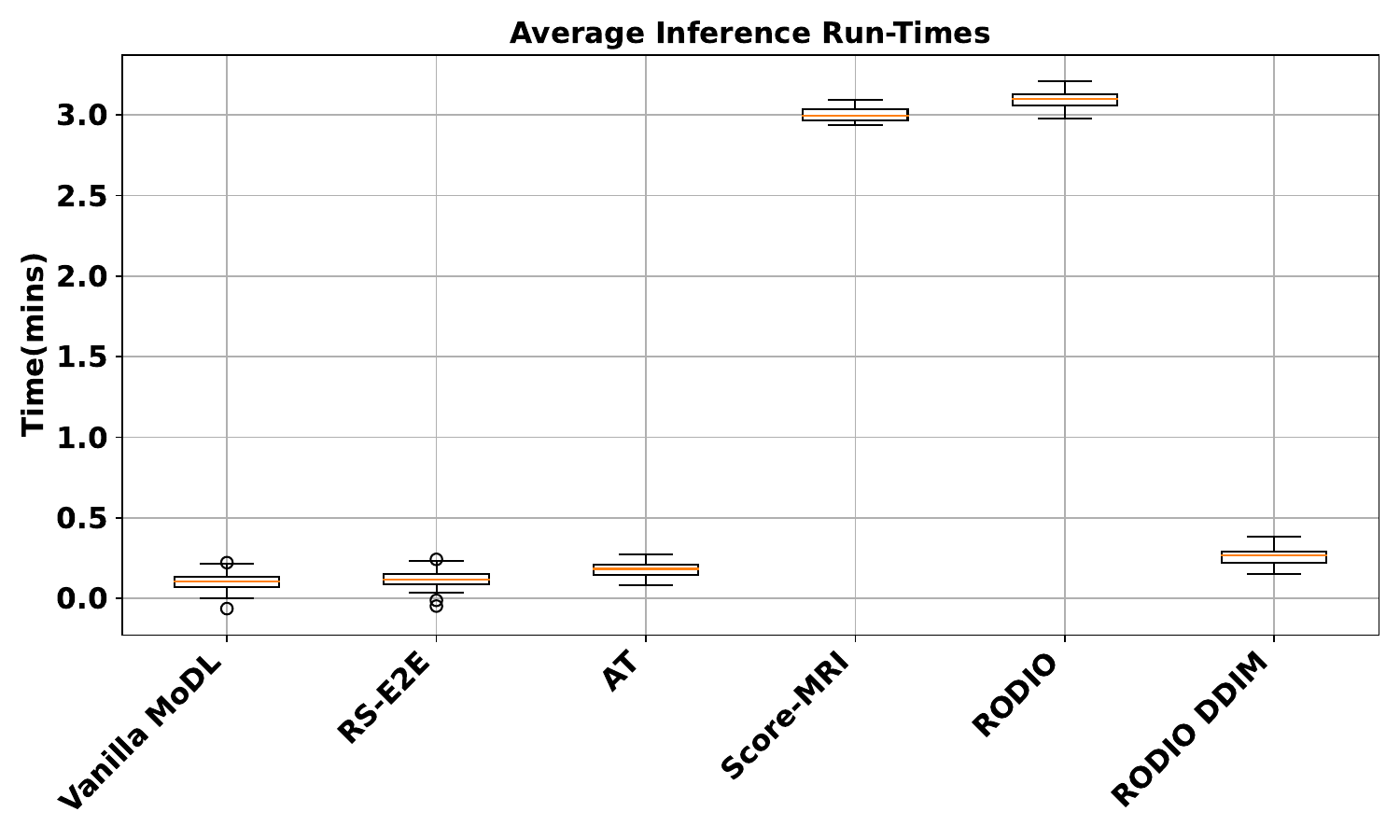}
\vspace{-0.3cm}
\caption{\small{\textcolor{black}{Average inference run-time of our proposed approach (\textcolor{black}{RODIO and RODIO-DDIM}) and the baselines for the experiment setting of the top right box plot of Figure~\ref{fig: box plot knee}.} }}
\label{fig: run time results}
\vspace{-0.7cm}
\end{figure}

\subsubsection{{Training/Testing Sampling Protocol and Undersampling Rate Disparities}}

\textcolor{black}{Here}, we \textcolor{black}{consider} two variations in the construction of the forward operator $\mathbf{A}$ between the training and testing phases. \textcolor{black}{In other words}, we train MoDL with $\mathbf{A}$ and evaluate it with different $\mathbf{A}_\textrm{test}$. The first variation involves using a different acceleration factor (sampling rate), while the second involves shifts in the locations of the k-space samples. \textcolor{black}{In particular, for the first variation, we train MoDL with 4x undersampling, and test it with \{2x,3x,4x,5x,6x,7x,8x\}. For the second variation, we train MoDL using a 4x mask and then evaluate it using various shifted versions of the original mask. Specifically, the central part of the mask (low frequencies) remains constant, whereas the higher frequency phase encodes are shifted by \{5\%,10\%,15\%,20\%,25\%\}.}  

\subsubsection{{Unseen Anatomies \& Pathologies at Testing Phase}}

\textcolor{black}{We evaluate our method's performance in the presence of white non-specific lesions using the fastMRI+ dataset\footnote{\tiny{\url{https://github.com/microsoft/fastmri-plus/tree/main}}}. In particular, the DL-based image reconstructor is trained on the lesion-free fastMRI dataset and evaluated on the fastMRI+ dataset}.

\textcolor{black}{Furthermore, we evaluate the performance of the proposed method with testing brain measurements but wherein the diffusion purifier was pre-trained on knee data (i.e., different anatomy).}



\subsection{\textcolor{black}{Studies on} the Process Switching Time (PST) Step}

In this section, we \textcolor{black}{first} conduct an experiment to evaluate the effectiveness of the proposed MMD-based method in determining the near-optimal PST step, denoted as $N_{t^*}$. The experiment is depicted in Figure~\ref{fig: tstar ablation} (\textit{top}), where we present the MMD values computed using \eqref{eqn: MMD} \textcolor{black}{for the fastMRI dataset}. Additionally, Figure~\ref{fig: tstar ablation} (\textit{bottom}) displays the results obtained when applying various values of $N_{t^*}$ within our pipeline, with corresponding PSNR values compared to ground truth images. 
In this experiment, we calculate the MMD values by setting the Gaussian kernel $v$ as the mean of the magnitude of the images in set $Z$, which comprises images $\mathbf{A}^H\mathbf{y}$ for 20 \textcolor{black}{knee fastMRI} scans $\mathbf{y}\in D$. For the perturbed images, we utilize the worst-case additive perturbations, denoted as $\bm{\delta}$, calculated from \eqref{eqn: attack}. Consequently, the set $Z_\textrm{p}$ encompasses $\mathbf{A}^H(\mathbf{y}+\bm{\delta})$ for the same measurements used in $Z$.

The results of Figure~\ref{fig: tstar ablation} (\textit{bottom}) show that, in comparison to the ground truth, the optimal PSNR result is achieved at $N_{t^*} = 150$, consistent with the observed approximate MMD value in Figure~\ref{fig: tstar ablation} (\textit{top}). Furthermore, it is evident that although the MMD values for $N_{t^*}$ in the range $(150,500]$ are also close to zero, PSNR values begin to deteriorate. This observation aligns with the intuition that increasing the value of $N_{t^*}$ effectively removes perturbations but runs the risk of losing image structure. Consequently, for the remainder of this paper, we adopt $N_{t^*} = 150$ as our chosen setting. We remark that the number of reverse (purification) process steps chosen for our robustification task, which is 150, is notably lower than the requirement in the diffusion-based image reconstruction task presented in \cite{chung2022score}, where 500 steps were used.

\textcolor{black}{In Figure~\ref{fig: tstar ablation 2}, we repeat the above experiment for another dataset, the Stanford 2D FSE Dataset \cite{FSE}, and we see similar observations.}



\textcolor{black}{We remark that we use the same PST for our other experiments that evaluate different instabilities. We generally observe that setting PST to 150 is sufficient. This (in addition to observing the same trend on another dataset) shows that this selection is general w.r.t. different instabilities and not highly sensitive. To support this, in Figure~\ref{fig: tstar ablation 3}, we report the results of our method using acceleration factors (3x and 8x) that are different from the one used during training (which is 4x) across different PST values. We note that this disparity between training and testing is an instability that is not an additive noise like the previous two results. As observed, the best results are obtained with PST set to 150. Furthermore, the results using different PSTs did not significantly degrade the performance as the the maximum drop in PSNR is less than 1.2 dB. }



\subsection{Robustness Results}\label{sec: exp rob results}

\begin{figure*}[t]
\centering
\includegraphics[width=\textwidth]{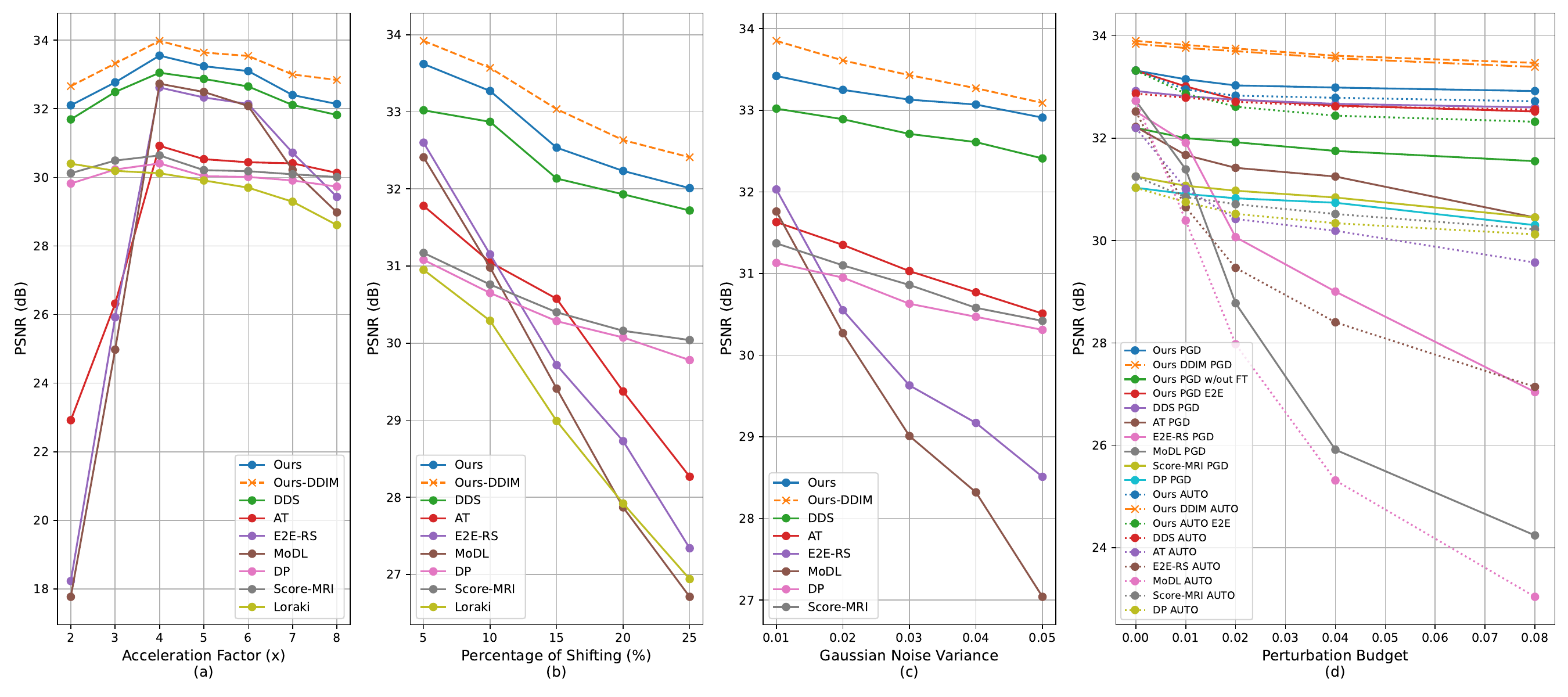}
\vspace{-0.88cm}
\caption{\small{Robustness evaluation against variations in: (a) acceleration factors, (b) locations of k-space sampling, (c) variance level of the Gaussian random additive noise, and (d) perturbation budget of the worst-case additive disturbances generated by PGD and AUTO methods. The `PGD E2E' and `AUTO E2E' \textcolor{black}{cases} in (d) correspond to the cases of generating end-to-end perturbations while calculating gradients through propagating the DP and MoDL. Furthermore, `Ours PGD w/out FT' corresponds to the case where no MoDL fine-tuning is applied. \textcolor{black}{`Ours with DDIM' corresponds to the case of running our algorithm with the DDIM accelerated sampler.} The figure is best viewed in color. }}
\label{fig: overall eval}
\end{figure*}
\begin{figure}[t]
\centering
\includegraphics[width=8.1cm]{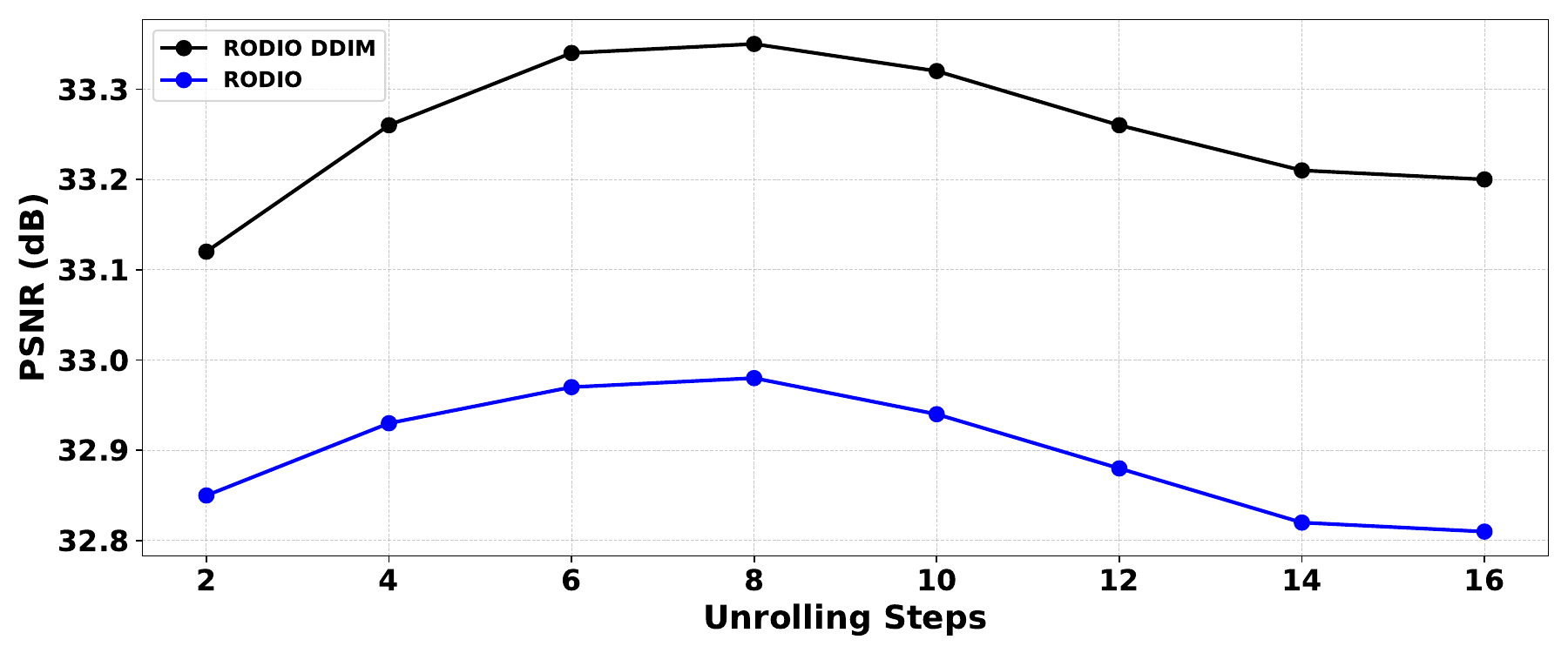}
\vspace{-0.3cm}
\caption{\small{\textcolor{black}{Average PSNR of our method (RODIO and RODIO-DDIM) using measurements with additive Gaussian noise (with 0.01 variance) across different unrolling steps (i.e., $N$ in MoDL).} }}
\label{fig: unrolling ablation with gaussian noise}
\vspace{-0.6cm}
\end{figure}
\begin{figure*}[!t]
\begin{tabular}[b]{cccccc}
    \textbf{Ground Truth}&
    \textbf{MoDL}&
    \textbf{RS-E2E}&
    \textbf{AT}&
    \textbf{Score-MRI}&
    \textbf{RODIO (Ours)}\\      \includegraphics[width=.14\linewidth,valign=t,angle=180]{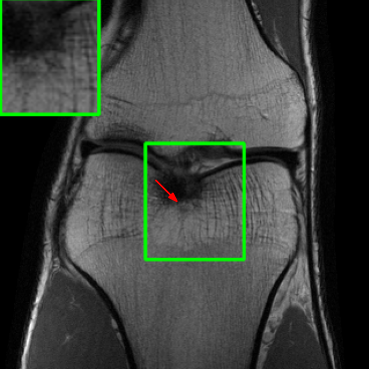}&
    \includegraphics[width=.14\linewidth,valign=t,angle=180]{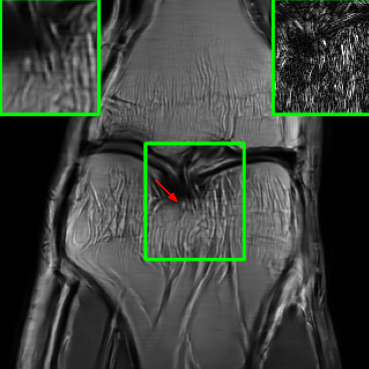}&
   \includegraphics[width=.14\linewidth,valign=t,angle=180]{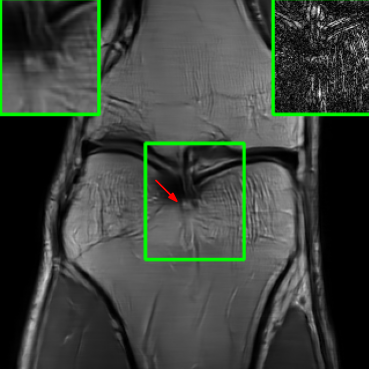} &
    \includegraphics[width=.14\linewidth,valign=t,angle=180]{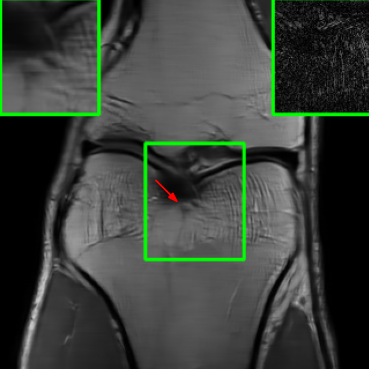}&
    \includegraphics[width=.14\linewidth,valign=t,angle=180]{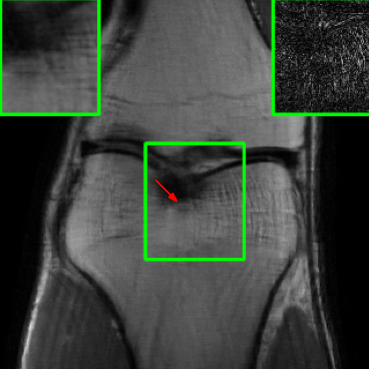}&
    \includegraphics[width=.14\linewidth,valign=t,angle=180]{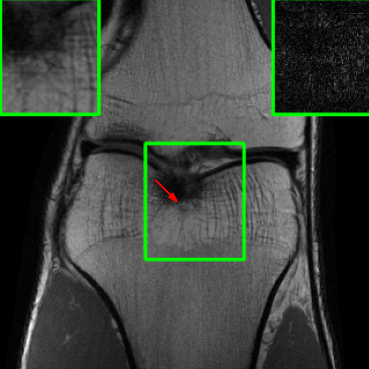}\\
    \scriptsize{PSNR = $\infty$  dB}  
   &\scriptsize{PSNR = 22.28 dB} 
    &\hspace{-0.2cm} 
    \scriptsize{PSNR = 25.34 dB}
    &\hspace{-0.2cm} 
     \scriptsize{PSNR = 29.47 dB}
    &\hspace{-0.2cm}
    \scriptsize{PSNR = 29.28 dB } 
    &\scriptsize{\textbf{PSNR = 32.88 dB} }\\
    \includegraphics[width=.14\linewidth,valign=t,angle=180]{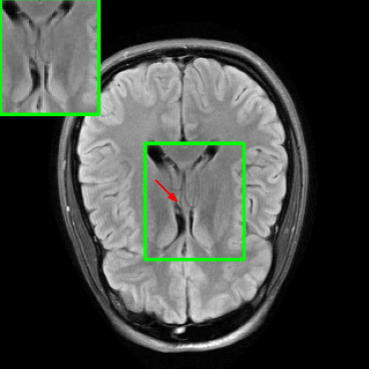}&
    \includegraphics[width=.14\linewidth,valign=t,angle=180]{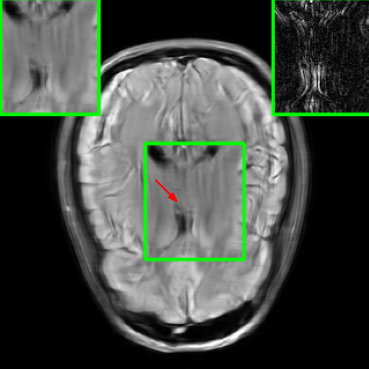}&
   \includegraphics[width=.14\linewidth,valign=t,angle=180]{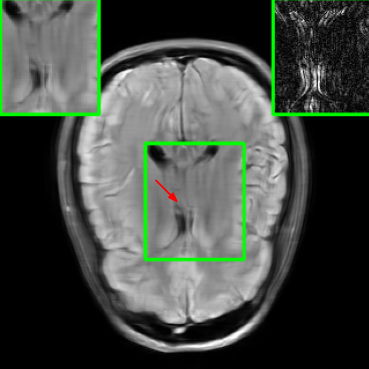} &
    \includegraphics[width=.14\linewidth,valign=t,angle=180]{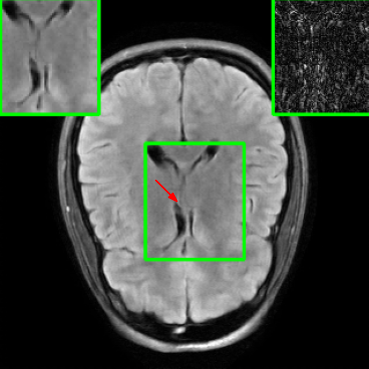}&
    \includegraphics[width=.14\linewidth,valign=t,angle=180]{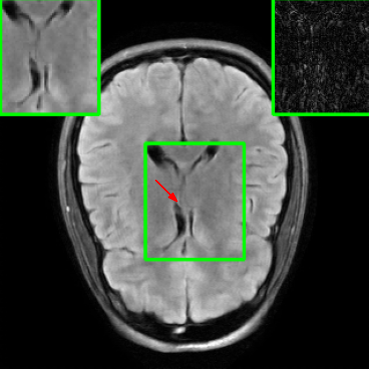}&
  \includegraphics[width=.14\linewidth,valign=t,angle=180]{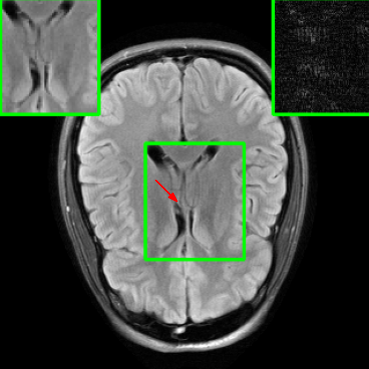}\\
    \scriptsize{PSNR = $\infty$  dB}  
   &\scriptsize{PSNR = 22.23 dB} 
    &\hspace{-0.2cm} 
    \scriptsize{PSNR = 24.56 dB}
    &\hspace{-0.2cm} 
    \scriptsize{PSNR = 29.25 dB }
      &\hspace{-0.2cm}
    \scriptsize{PSNR = 29.35 dB}
    &\hspace{-0.2cm}
    \scriptsize{\textbf{PSNR = 33.18 dB} }
        \\ 
    
\end{tabular}
\vspace{-0.3 cm}
\caption{\small{\textcolor{black}{Visualization of ground-truth and reconstructed images using different methods, evaluated by PGD-based worst-case additive perturbations with $\epsilon = 0.02$.}}}
\label{fig:denoised_imgs_zoomed_300case}
\vspace{-0.2 in}
\end{figure*}
\begin{figure*}[!t]
\begin{tabular}[b]{cccccc}
    \textbf{Ground Truth}&
    \textbf{MoDL}&
    \textbf{RS-E2E}&
    \textbf{AT}&
    \textbf{Score-MRI}&
    \textbf{RODIO (Ours)}\\      \includegraphics[width=.14\linewidth,valign=t]{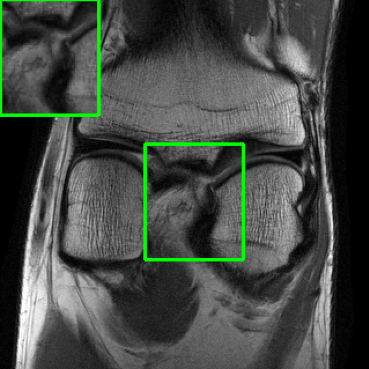}&
    \includegraphics[width=.14\linewidth,valign=t]{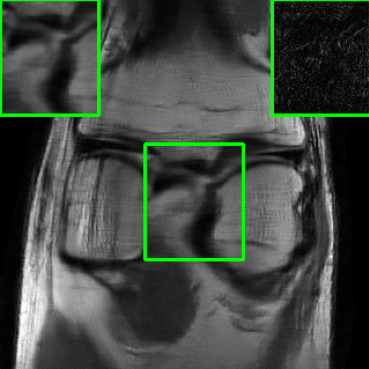}&
   \includegraphics[width=.14\linewidth,valign=t]{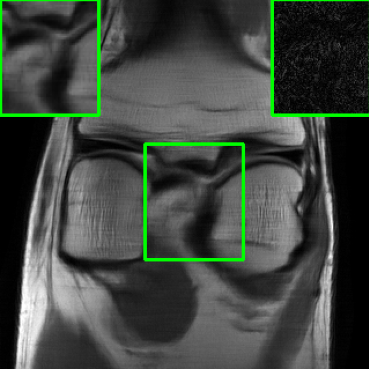} &
    \includegraphics[width=.14\linewidth,valign=t]{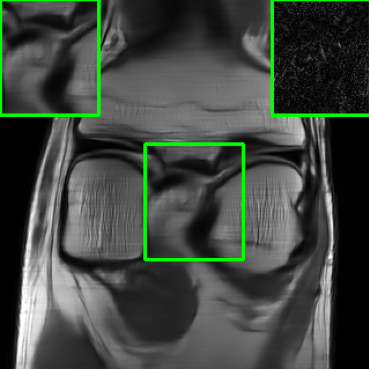}&
    \includegraphics[width=.14\linewidth,valign=t]{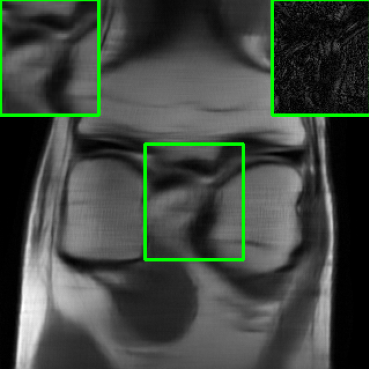}&
    \includegraphics[width=.14\linewidth,valign=t]{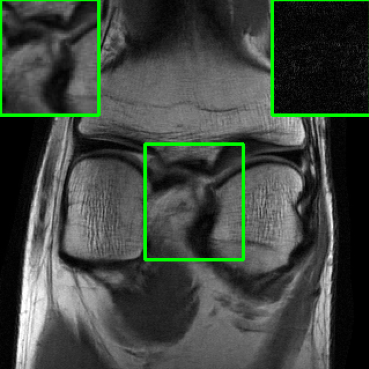}\\
    \scriptsize{PSNR = $\infty$  dB}  
   &\scriptsize{PSNR = 32.28 dB} 
    &\hspace{-0.2cm} 
    \scriptsize{PSNR = 31.13 dB}
    &\hspace{-0.2cm} 
     \scriptsize{PSNR = 31.07 dB}
    &\hspace{-0.2cm}
    \scriptsize{PSNR = 30.87 dB } 
    &\scriptsize{\textbf{PSNR = 32.67 dB} }
\end{tabular}
\vspace{-0.3 cm}
\caption{ \small{\textcolor{black}{Visualization of ground-truth and reconstructed images using different methods, evaluated by the knee fastMRI testing set with 8x acceleration factor.}}}
\label{fig:denoised_imgs_zoomed_8x}
\vspace{-0.18 in}
\end{figure*}

\begin{figure*}
    \centering
    \includegraphics[width=0.81\textwidth]{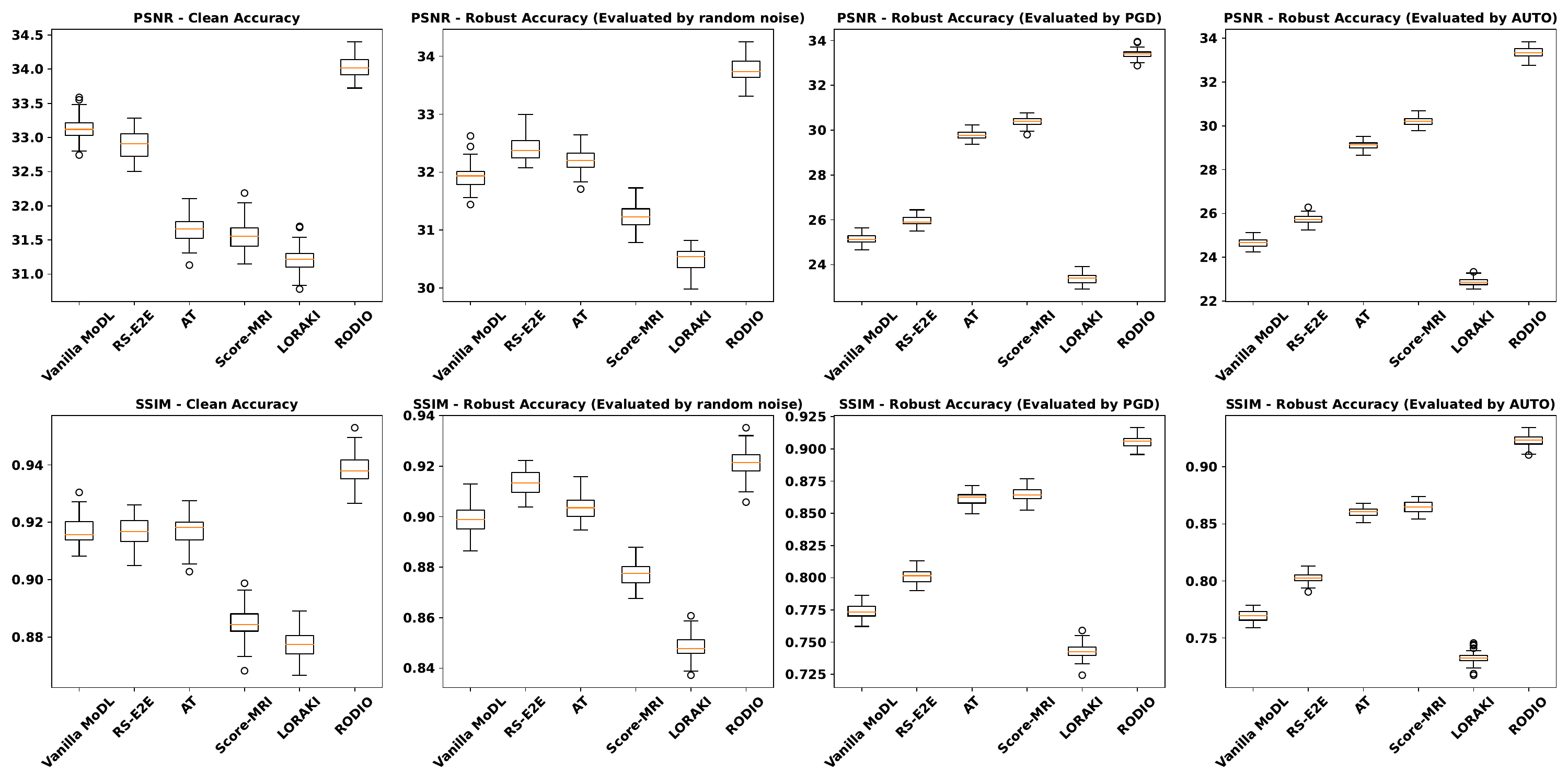}
    \vspace{-0.35cm}
    \caption{\small{ \textcolor{black}{Reconstruction accuracy box plots of the \textbf{brain} 
fastMRI dataset with 4x Acceleration factor.} The additive Gaussian random noise of the second column plots are obtained from using variance of $0.01$. The worst-case additive noise of the third and fourth columns are obtained using PGD and AUTO methods with $\epsilon = 0.02$. }}
    \label{fig: box plot brain}
    \vspace{-0.1 in}
\end{figure*}

\begin{figure*}[!t]
\begin{tabular}[b]{cccccc}
    \textbf{Ground Truth}&
    \textbf{MoDL}&
    \textbf{RS-E2E}&
    \textbf{AT}&
    \textbf{Score-MRI}&
    \textbf{RODIO (Ours)}\\      \includegraphics[width=.14\linewidth,valign=t,angle=180]{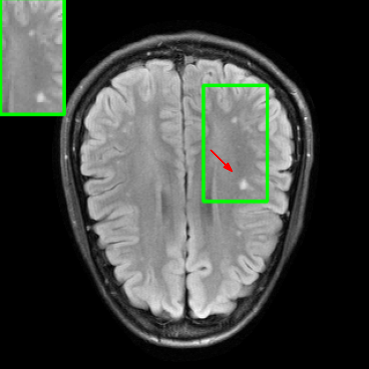}&
    \includegraphics[width=.14\linewidth,valign=t,angle=180]{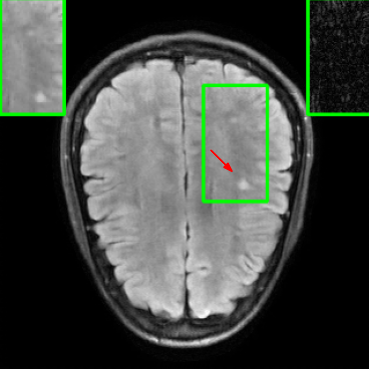}&
   \includegraphics[width=.14\linewidth,valign=t,angle=180]{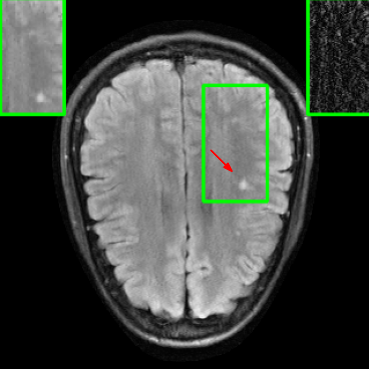} &
    \includegraphics[width=.14\linewidth,valign=t,angle=180]{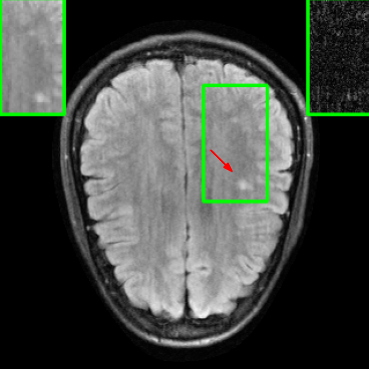}&
    \includegraphics[width=.14\linewidth,valign=t,angle=180]{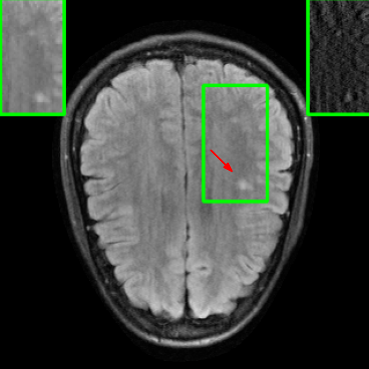}&
    \includegraphics[width=.14\linewidth,valign=t,angle=180]{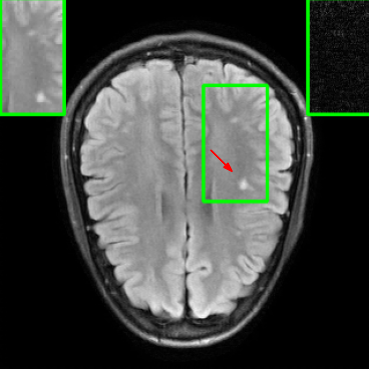}\\
    \scriptsize{PSNR = $\infty$  dB}  
   &\scriptsize{PSNR = 34.28 dB} 
    &\hspace{-0.2cm} 
    \scriptsize{PSNR = 33.13 dB}
    &\hspace{-0.2cm} 
     \scriptsize{PSNR = 34.07 dB}
    &\hspace{-0.2cm}
    \scriptsize{PSNR = 32.27 dB } 
    &\scriptsize{\textbf{PSNR = 34.92 dB} }
\end{tabular}
\vspace{-0.3 cm}
\caption{\small{\textcolor{black}{Visualization of ground-truth and reconstructed images using different methods, \textcolor{black}{trained with fastMRI (without lesions), and evaluated by the fastMRI+ dataset (with lesions)}.}}}
\label{fig:denoised_imgs_zoomed_lession}
\vspace{-0.0 in}
\end{figure*}

\subsubsection{Robustness to Additive Perturbations}

Figure~\ref{fig: box plot knee} presents box plots 
for a comprehensive view of the performance of our robustification method (\textcolor{black}{with the sampler in Algorithm~\ref{alg: PC with DC} which we term by RODIO and with the DDIM accelerated sampler which is labeled as RODIO-DDIM}), as well as that of \textcolor{black}{Vanilla MoDL}, AT, E2E-RS, \textcolor{black}{LORAKI}, Score-MRI, \textcolor{black}{and DDS} assessed through PSNR (top) and SSIM (bottom) metrics using the knee dataset. We evaluate these methods across multiple scenarios, including benign aliased images (top and bottom first plots), images subjected to additive random Gaussian noise with variance of $0.01$ (top and bottom second plots), and images with additive worst-case perturbations generated using PGD and AUTO methods with $\epsilon = 0.02$ (top and bottom last two plots).

While AT, RS, \textcolor{black}{DDS}, and score-MRI show improvements when compared to vanilla MoDL, we observe that, on average, our robustification approach reports the highest values of PSNR and SSIM. For the example of the rightmost plot, our method achieves an average PSNR that is approximately 3 dB more than score-MRI and nearly 9 dB more than Vanilla MoDL. Additionally, the PSNR and SSIM results in the first plots (top and bottom) indicate an improvement of our proposed approach (\textcolor{black}{RODIO and RODIO-DDIM}), even in the absence of any perturbations. 





Although our proposed approach \textcolor{black}{(RODIO and RODIO-DDIM)} report the highest PSNR values in terms of reconstruction, \textcolor{black}{RODIO} requires larger inference run-time compared to AT, RS, vanilla MoDL, and \textcolor{black}{RODIO-DDIM}. In Figure~\ref{fig: run time results}, we present inference run-times for the setting of the top right box plot of Figure~\ref{fig: box plot knee}. As observed, on average, our method (\textcolor{black}{combined with Algorithm~\ref{alg: PC with DC}}) and score-MRI need nearly 3 minutes per image, whereas \textcolor{black}{RODIO-DDIM and the} other methods require only 60 seconds or less. The increased run-time \textcolor{black}{in RODIO} is attributed to the application of the proposed diffusion purification prior to DL-reconstructor, representing a trade-off. \textcolor{black}{The results of RODIO-DDIM show that our proposed approach can be adapted to accelerated samplers with a run-time that is nearly close to inference time of other robustification approaches such as AT.}

In Figure~\ref{fig: overall eval}~(c), we present the PSNR values of AT, E2E-RS, DP, score-MRI, \textcolor{black}{DDS}, and our approach (\textcolor{black}{RODIO and RODIO-DDIM}), evaluated under different levels of added Gaussian \textcolor{black}{measurements} noise during testing. Notably, as the noise level (indicated by the variance) increases, the reported PSNR values decrease for all methods. However, our approach consistently reports higher PSNR values when compared to the other baselines across all tested noise levels. For instance, when faced with a variance of $0.05$, our methods (\textcolor{black}{RODIO}) reports nearly 33 dB whereas the second best \textcolor{black}{robustification method} (in this case AT) reports a PSNR of 30.5 dB.

\textcolor{black}{In Figure~\ref{fig: unrolling ablation with gaussian noise}, we report the results of our methods (RODIO and RODIO-DDIM) across different unrolling steps (i.e., $N$ in Algorithm~\ref{alg: Robust MoDL Pipeline}) in the x-axis with measurement Gaussian noise of variance $0.01$. As observed, our methods consistently return high PSNR values even when the unrolling steps are below or above 8 steps (the setting in our experiments). This demonstrates that RODIO is not sensitive to non-DP hyper-parameters.}

In Figure~\ref{fig: overall eval}~(d), we present the PSNR performance of our approach and the considered baselines, evaluated under varying perturbation budgets given by the values of $\epsilon$. \textcolor{black}{For the worst-case perturbations,} the evaluation includes both PGD and AUTO methods. Additionally, we include the PGD E2E and AUTO E2E scenarios, which involve generating end-to-end perturbations using \eqref{eqn: attack e2e}. \textcolor{black}{Moreover, the results of our method with the DDIM accelerated sampler are included (labeled as ``Ours DDIM PGD'' and ``Ours DDIM AUTO'')}. As the perturbation budget increases, all methods experience a decline in their PSNR values, which is expected. However, we observe that our approach consistently returns the highest PSNR values across the entire range of perturbation budgets. We also observe that employing the E2E attack results in slightly lower PSNR values compared to the case of generating perturbations solely w.r.t. MoDL. Finally, we observe that the AUTO results are marginally lower than those of PGD, which aligns with expectations since AUTO represents a more advanced approach in generating worst-case additive noise. 

Moreover, in Figure~\ref{fig: overall eval}~(d), we illustrate the effect of fine-tuning on the robustness of our method. Specifically, we compare PSNR values for our approach when exposed to PGD-based worst-case additive perturbations under two scenarios: with fine-tuning MoDL using perturbed purified training samples (i.e., $f_{\theta_{\textrm{FT}}}$) and without fine-tuning, relying solely on the pre-trained MoDL (i.e., $f_\theta$). These two cases are represented by the solid black and solid green \textcolor{black}{curves} in Figure~\ref{fig: overall eval}~(d). The results clearly highlight that the pre-trained+fine-tuned MoDL enhances robustness, as evidenced by the higher PSNR values compared to pre-trained MoDL. We also note that the results obtained without fine-tuning are slightly higher than those achieved using AT (see the solid brown curve in Figure~\ref{fig: overall eval}~(d)). This indicates that MoDL+DP without fine-tuning still exhibits improvements when compared to AT, vanilla MoDL, and RS-E2E. \textcolor{black}{We note that the results of DDS in Figure~\ref{fig: overall eval}~(c) and in Figure~\ref{fig: overall eval}~(d) indicate that this method is more robust than many other approaches. However, our method with DDIM still outperform DDS by nearly 1dB. We emphasize that DDS is a diffusion-based method that starts from random noise, whereas our method is a robustification method for supervised DL-based reconstructors that uses diffusion models as purifiers.}

 Figure~\ref{fig:denoised_imgs_zoomed_300case} presents visual comparison of image reconstructions and their associated reconstruction errors within a closely examined region. Each image in the figure includes two inset panels in the bottom-left and bottom-right corners. The bottom-left inset panel, enclosed within a green bounding box, serves as a reference for the region of interest in the image. In contrast, the bottom-right inset panel depicts an error map in relation to the ground truth. Notably, our method stands out in its ability to capture \textcolor{black}{more features} from the original image, surpassing the performance of alternative methods (as also evident from the reported PSNR values). 

\subsubsection{Robustness to Different Sampling Protocols \& Undersampling Rates}

In Figure~\ref{fig: overall eval}~(a), we illustrate the performance across different acceleration factors. During training, a k-space undersampling or acceleration factor of 4x was employed. However, during testing, we assess performance with various acceleration factors ranging from 2x to 8x. It is evident that when the acceleration factor matches the training phase (4x), all methods exhibit their highest PSNR results compared to when different acceleration factors are used. Nevertheless, when compared to the other methods, our approaches (\textcolor{black}{labeled in the legend as `Ours' and `Ours-DDIM'}) consistently \textcolor{black}{report} the highest PSNR values when tested with acceleration factors other than 4x. For instance, at 2x acceleration, \textcolor{black}{other robustification methods} (AT and E2E-RS) report PSNR values of 23 dB or lower, while our approach achieves nearly 32 dB. Additionally, in Figure~\ref{fig: overall eval} (a), we report results of using LORAKI with different acceleration factors. As observed, LORAKI \textcolor{black}{reports lower PSNR values} when compared to our proposed approach.

Figure~\ref{fig: overall eval}~(b) shows the PSNR values of our proposed approach and the considered baselines, assessed under varying percentages of shifts in the location of the k-space sampling during testing. The shifts were applied to high-frequency phase encode locations in the original sampling pattern or mask. This is to help understand reconstruction robustness when the sampling masks change a lot at a fixed k-space undersampling factor. We observe that as the percentage of shifts increases, the reported PSNR values decrease across all methods. However, we observe that, our method consistently outperforms the other approaches across all tested percentages, exhibiting the highest PSNR values. For instance, when the mask at testing time contains 25\% shift when compared to the training mask, our methods achieve \textcolor{black}{approximately 32 dB and 32.5 dB} whereas all other \textcolor{black}{robustification} methods report PSNR values of nearly 30 dB or less. \textcolor{black}{The results of DDS are also more robust than other methods. However, our method (with PCDC and DDIM) shows better results.} 
\begin{table}[!t]
\centering
\caption{\small{\textcolor{black}{Reconstruction accuracy for fastMRI knee data using the testing portion of the dataset with acceleration of \textbf{8x}.  }}}
\label{tab: 4x 8x results}
\vspace*{-1mm}
\resizebox{0.43\textwidth}{!}{%
\begin{tabular}{c|cc|c}
\toprule[1pt]
\midrule
Models 
& \multicolumn{2}{c}{Accuracy}  
& Training 
\\
Metrics 
& PSNR $\uparrow$ & SSIM $\uparrow$ & Acceleration Factor
\\
\midrule
Vanilla MoDL
& 33.25
& 0.920
& 8x
\\
E2E-RS
& 33.12
& 0.917
& 8x
\\
AT
& 32.17
& 0.913
& 8x
\\
Score-MRI
& 33.5
& 0.899
& 4x
\\
\midrule
\textcolor{black}{RODIO} (DP+MoDL)
& 33.67
& 0.922
& 4x
\\
\midrule
\bottomrule[1pt]
\end{tabular}}
\vspace*{-2.5mm}
\end{table}

\textcolor{black}{To further underscore the generalization and robustness of our proposed approach, we designed an experiment with \textcolor{black}{different} training and testing settings across different methods. Specifically, we trained vanilla MoDL, AT, and RS models using an 8x acceleration factor, while our method and score-MRI were trained with a 4x acceleration factor. Subsequently, we subjected benign measurements to testing with an 8x acceleration factor, aligning with the training settings of MoDL, AT, and RS, rather than 4x. The results, \textcolor{black}{given} in Table~\ref{tab: 4x 8x results}, showcase that our method, despite undergoing testing with a \textcolor{black}{different} acceleration setting, \textcolor{black}{reports slightly higher PSNR (33.67 dB) and SSIM (0.922) values when compared to other methods}. Moreover, the visualizations in Figure~\ref{fig:denoised_imgs_zoomed_8x} \textcolor{black}{show that} when tested with an 8x acceleration factor despite being trained on 4x, our proposed approach outperforms \textcolor{black}{the considered baselines} under conditions where both training and testing acceleration factors are 8x.}

\subsubsection{Robustness to Anatomical Variations}

\begin{table}[!t]
\centering
\caption{\small{\textcolor{black}{Brain fastMRI+ (with lesion) results. }}}
\label{tab: Lesion Brain}
\vspace*{-1mm}
\resizebox{0.35\textwidth}{!}{%
\begin{tabular}{c|cc}
\toprule[1pt]
\midrule
Models 
& \multicolumn{2}{c}{MRI Reconstruction Accuracy}  
\\
Metrics 
& PSNR $\uparrow$ & SSIM $\uparrow$
\\
\midrule
Vanilla MoDL
& 31.25
& 0.915
\\
E2E-RS
& 31.12
& 0.912
\\
AT
& 30.87
& 0.910
\\
Score-MRI
& 30.22
& 0.885
\\
\midrule
\textcolor{black}{RODIO} (DP+MoDL)
& 32.4
& 0.919
\\
\midrule
\bottomrule[1pt]
\end{tabular}}
\vspace*{-4.5mm}
\end{table}
\begin{figure}[t]
\centering
\includegraphics[width=7.0cm]{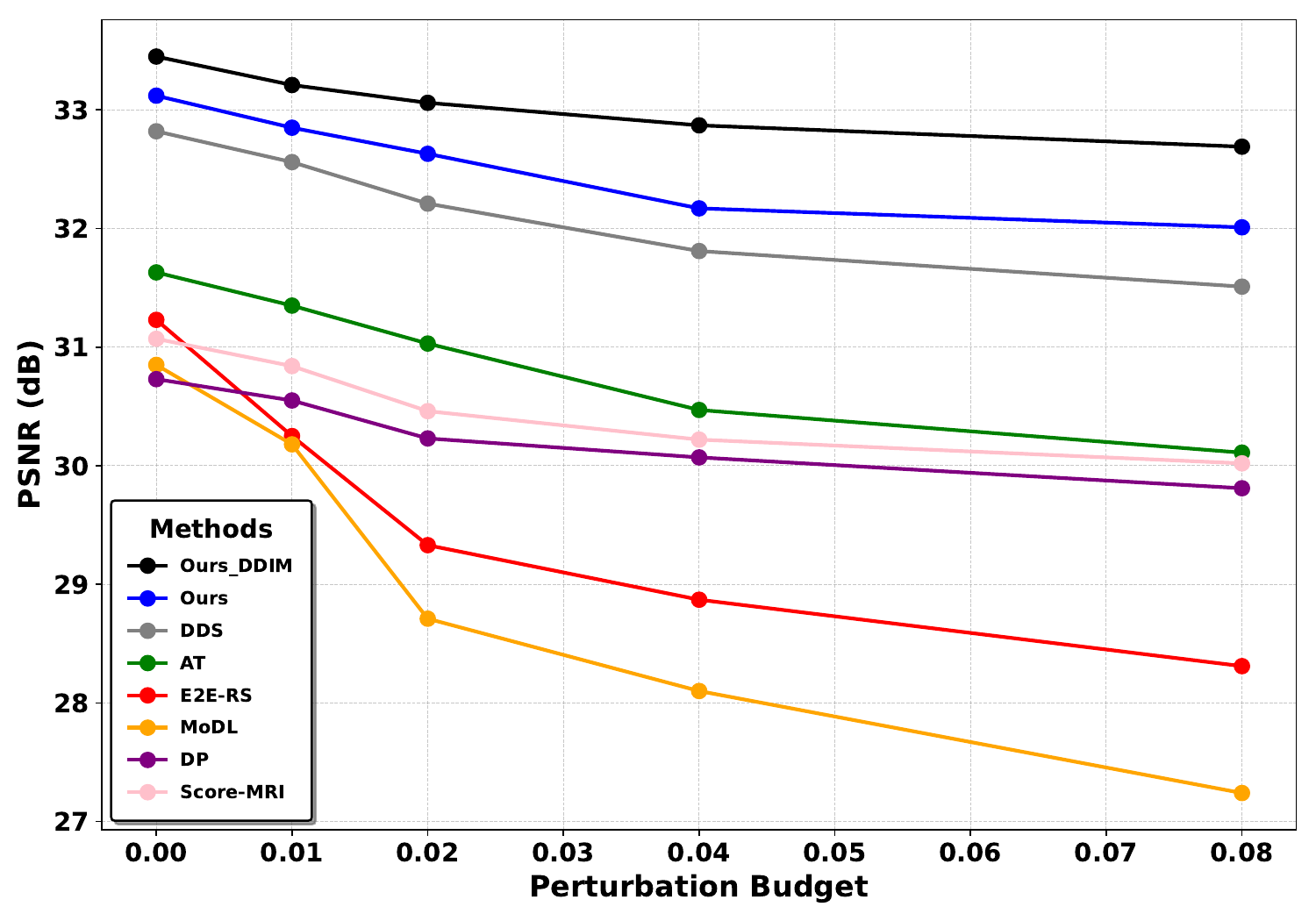}
\vspace{-0.3cm}
\caption{\small{\textcolor{black}{Reconstruction performance (PSNRs in y-axis) of our method and baselines using measurements with two sources of instabilities: unseen lesions and worst-case additive noise with perturbation budgets indicated on the x-axis.}}}
\label{fig: combined lesions and pertubations}
\vspace{-0.6cm}
\end{figure}
\begin{table*}[t]
\centering
\vspace*{-3mm}
\caption{\small{\textcolor{black}{\textbf{Brain} dataset reconstruction accuracy using \textbf{Recurrent Variational Network} as our DL-based image reconstructor. }} }
\label{tab: brain varRecurrentNet}
\vspace*{-2mm}
\resizebox{0.9\textwidth}{!}{%
\begin{tabular}{c|cc|cc|cc}
\toprule[1pt]
\midrule
Models 
& \multicolumn{2}{c}{Clean Accuracy}  
& \multicolumn{2}{c}{Robust Accuracy (Evaluated by random noise)} 
& \multicolumn{2}{c}{Robust Accuracy (Evaluated by PGD)} 
\\
Metrics 
& PSNR $\uparrow$ & SSIM $\uparrow$
& PSNR $\uparrow$ & SSIM $\uparrow$
& PSNR $\uparrow$ & SSIM $\uparrow$
\\
\midrule
Vanilla RecurrentVarNet
& 33.78
& 0.925
& 32.89
& 0.91
& 26.5
& 0.793
\\
AT+RecurrentVarNet 
& 33.19
& 0.919
& 33.01
& 0.914
& 31.67
& 0.892
\\
E2E-RS+RecurrentVarNet
& 33.67
& 0.922
& 33.12
& 0.915
& 30.20
& 0.875
\\
\midrule
\textcolor{black}{RODIO} (DP+RecurrentVarNet) (Ours)
& 34.33
& 0.941
& 34.07
& 0.938
& 33.64
& 0.935
\\
\midrule
\bottomrule[1pt]
\end{tabular}}
\vspace*{-4.5mm}
\end{table*}

\begin{figure*}[t]

\begin{tabular}[b]{ccccc}
    \textbf{Ground Truth}&
    \textbf{RecurrentVarNet}&
    \textbf{E2E-RS}&
    \textbf{AT}&
    \textbf{RODIO (Ours)}\\
        \includegraphics[width=.17\linewidth, valign=t,angle=180]{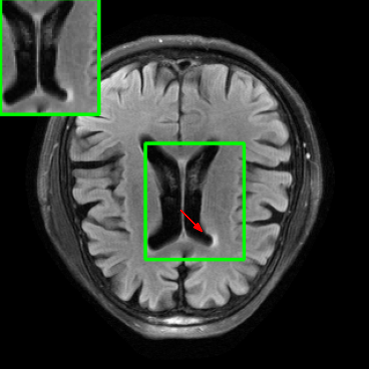}
        &
        \includegraphics[width=.17\linewidth, valign=t,angle=180]{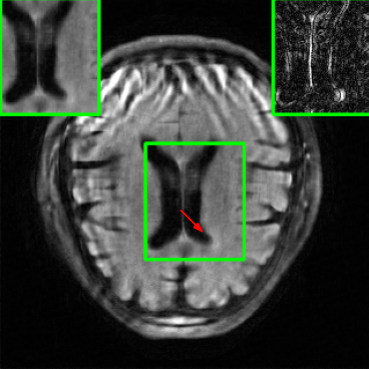}
        &
        \includegraphics[width=.17\linewidth, valign=t,angle=180]{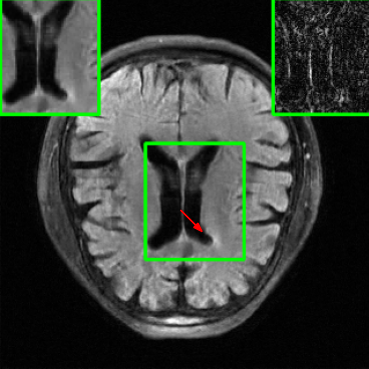}
        &
        \includegraphics[width=.17\linewidth, valign=t,angle=180]{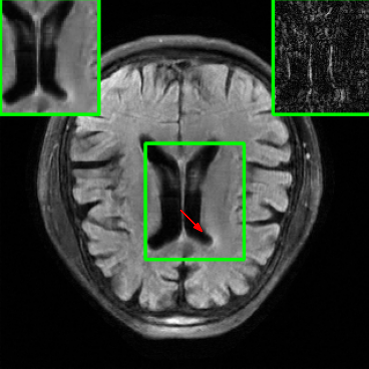}
        &
        \includegraphics[width=.17\linewidth, valign=t,angle=180]{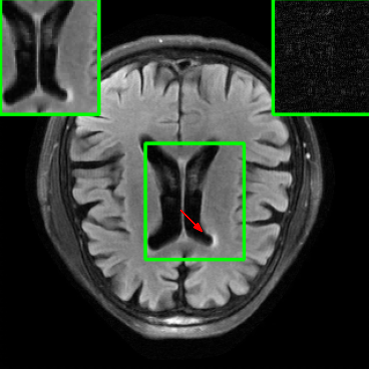}
        \\
    \scriptsize{PSNR = $\infty$  dB}  
   &\scriptsize{PSNR = 29.82 dB} 
   &\scriptsize{PSNR = 32.82 dB} 
    &\scriptsize{PSNR = 33.28 dB}
   & \scriptsize{PSNR = 35.78 dB}\\
\end{tabular}
\vspace{-.3cm}
\caption{\small{\textcolor{black}{Visualization of ground-truth and reconstructed images using RecurrentVarNet and RecurrentVarNet+DP (Ours) methods, evaluated by PGD-based worst-case additive perturbations with $\epsilon = 0.02$.}} 
}
\label{fig: RecurrentVarNet Vis}
\vspace*{-6mm}
\end{figure*}

In Figure~\ref{fig: box plot brain}, we replicate the experiment conducted in Figure~\ref{fig: box plot knee}, this time utilizing the brain dataset. Notably, MoDL underwent fine-tuning using perturbed purified examples sourced from the training set of the brain dataset. When comparing the results of our proposed method with other approaches, we find that the observations of Figure~\ref{fig: box plot knee} remain consistent. \textcolor{black}{For the PGD case (third column), our method reports an average SSIM of nearly 0.91 whereas Vanilla MoDL (the DL-reconstructor considered in this experiment) reports an average SSIM of approximately 0.775.} An important point to highlight is that the pre-trained DM employed in our purification stage for this experiment was originally trained exclusively on knee data, without any exposure to brain data. This underscores the robust generalization capabilities of the diffusion purification process within our approach, extending its effectiveness to previously unseen MRI datasets. 


We note that similar diffusion model generalization capabilities were also observed in \cite{chung2022score}. However, further thorough investigation is required to precisely determine the limitations of these generalization capabilities, and this remains a promising direction for future research.

Here, we employ the fastMRI+ dataset to assess our approach's image reconstruction capability. For the training phase, we employ the original fastMRI brain dataset, which excludes lesion cases, as the basis for training all methods. During the testing phase, however, we utilize the lesion dataset. Table~\ref{tab: Lesion Brain} shows the results, where our method reports the highest PSNR and SSIM values compared to other baselines. It is important to highlight that, unlike the cases of additive k-space noise and training/testing sampling protocol and undersampling rate disparities, the improvements observed from utilizing our method with unseen lesions are somewhat marginal as seen from the average PSNR and SSIM results (at least 1.2 dB PSNR improvement when compared to the 2nd best results). 
Additionally, visualizations are provided Figure~\ref{fig:denoised_imgs_zoomed_lession} where we highlight the nonspecific white matter lesion area. As observed, both visually and in terms of PSNR values, our approach reports improved results when compared to the other baselines.

\textcolor{black}{In Figure~\ref{fig: combined lesions and pertubations}, we report results (PSNR in the y-axis) of our method (labeled as `Ours' and `Ours DDIM') and baselines using measurements with unseen lesions combined with worst-case additive noise (generated by the PGD method) with budget indicated in the x-axis. We include these results to evaluate our method and baselines using a scenario that combines multiple sources of instabilities. As observed, our methods report the best results. }

\subsection{Applying \textcolor{black}{RODIO} to Other DL-based MRI Reconstruction Models}
\label{sec: supp other than modl}

\textcolor{black}{Here, we demonstrate the applicability of our diffusion purification strategy to other DL-based supervised MRI reconstructors. Specifically, we explore the Recurrent Variational Network (RecurrentVarNet) \cite{yiasemis2022recurrent}, presenting results both with and without perturbations, as well as with and without the integration of our diffusion purification technique. The results are summarized in Table~\ref{tab: brain varRecurrentNet}. As depicted in the table, when the standalone RecurrentVarNet (or RecurrentVarNet integrated with AT and/or RS) encounters additive worst-case perturbations in the measurement space, the reported PSNR and SSIM scores (last two columns of the first three rows) experience a significant drop (\textcolor{black}{for example, the Vanilla RecurrentVarNet encounters a PSNR drop of nearly 7 dB}). However, upon employing our diffusion purification (last row), we observe only a marginal decrease in performance (\textcolor{black}{of 0.69 dB}). These findings illustrate that our strategy can be integrated well with general DL-based reconstructors. The visualizations in Figure~\ref{fig: RecurrentVarNet Vis} provide additional support for our claim.}


\subsection{\textcolor{black}{Applying RODIO to Limited-data Unsupervised Setting}}
\label{sec: DP with unsupervised}

\begin{figure}[t]
\centering
\includegraphics[width=6.6cm]{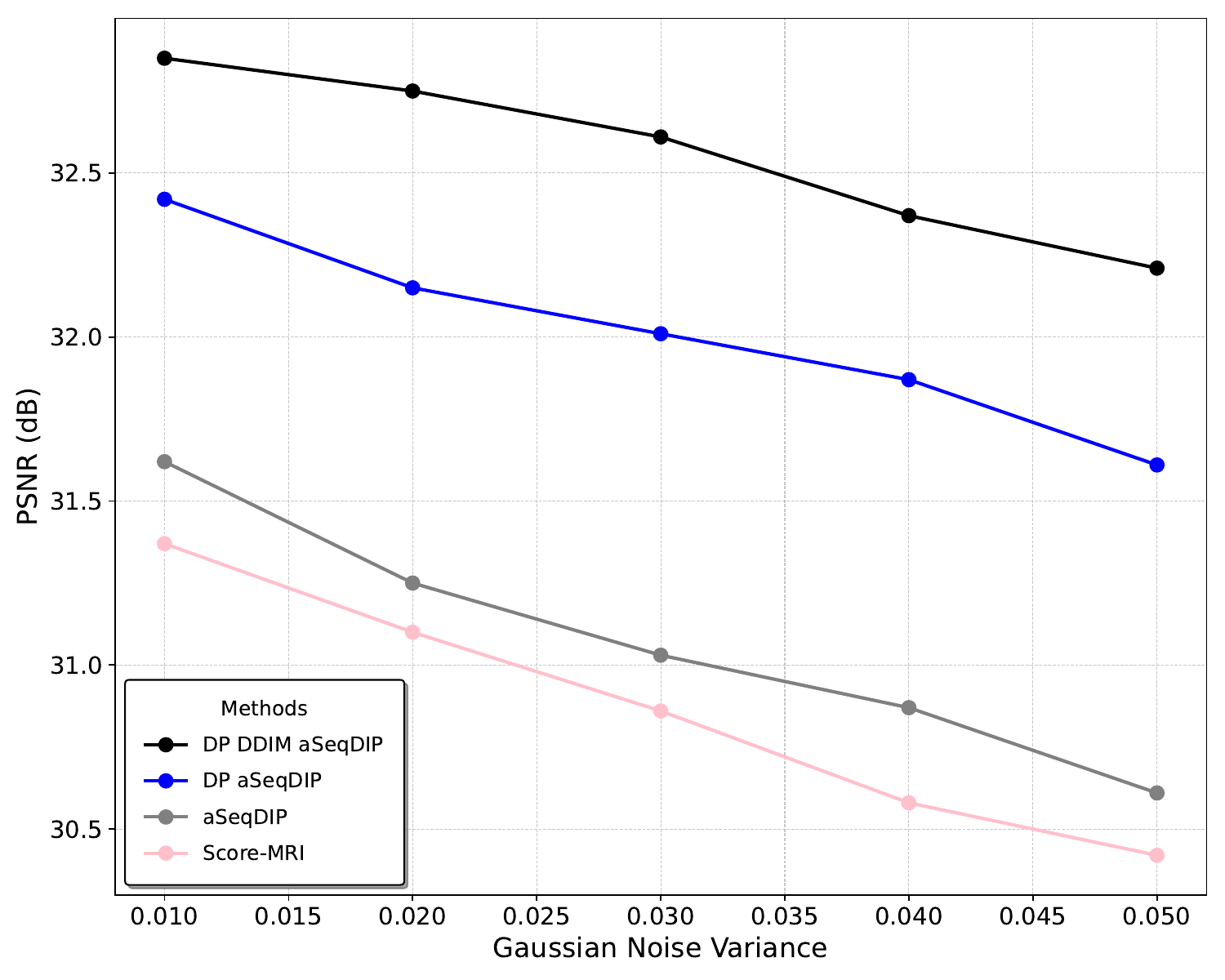}
\vspace{-0.3cm}
\caption{\small{\textcolor{black}{Average PSNR of aSeqDIP \cite{alkhouri2024image} (an unsupervised DL-based MRI reconstructor) with DP (representing RODIO) and without DP, using measurements with Gaussian noise with variance indicated in the x-axis.}}}
\label{fig: aSeqDIP vs. aSeqDIP+DP}
\vspace{-0.6cm}
\end{figure}

\textcolor{black}{In this subsection, we conduct an experiment to demonstrate that our diffusion purification (DP) approach can enhance the robustness of unsupervised DL-based reconstruction methods in limited-data settings. Specifically, we consider a recent DIP-based model that operates without training data \cite{alkhouri2024image}, termed auto-encoding sequential DIP (aSeqDIP). In Figure~\ref{fig: aSeqDIP vs. aSeqDIP+DP}, we report the average PSNR (y-axis) of aSeqDIP (gray curve) and RODIO-aSeqDIP (black and black curves using Algorithm~\ref{alg: PC with DC} and DDIM, respectively) when reconstructing measurements subject to additive Gaussian noise with varying variances (x-axis). As observed, RODIO-aSeqDIP achieves higher PSNR values, indicating that DP improves the robustness of aSeqDIP in the presence of additive measurement noise.} 

\section{Conclusion \textcolor{black}{\& Future work}}
\label{sec: conc}
Recent studies have unmasked vulnerabilities in DL-based MRI reconstruction methods—namely, susceptibility to 
additive perturbations and variations in training/testing settings, such as acceleration factors and k-space sampling patterns. This paper has addressed these 
challenges by harnessing the power of diffusion models. Our robustification strategy enhanced the resilience of DL-based MRI reconstruction models by integrating pre-trained diffusion models as noise purifiers. Unlike conventional robustification techniques like adversarial training (AT), our method eliminated the need for complex minimax optimization problems. Instead, it simply requires fine-tuning on perturbed purified examples. Our extensive experiments have illustrated the remarkable efficacy of our approach in mitigating different instabilities when compared to utilizig diffusion-based MRI reconstructors and leading robustification methods, including AT and randomized smoothing. We also evaluated the robustness of our approach using an MRI dataset with lesions. Moreover, we illustrated the adaptability of our strategy to multiple reconstruction models. These findings underscore the promise of leveraging diffusion models to enhance the robustness and reliability of DL-based MRI reconstruction, paving the way for more dependable and accurate medical imaging technologies in the future.


\textcolor{black}{For future research, we aim to explore other modalities such as medical computed tomography and non-medical image recovery tasks. Furthermore, recent studies in low-level vision tasks (e.g., \cite{wang2023gridformer,zhang2021ddmsnet}) demonstrate the strength of multiscale designs for robust feature extraction under various perturbations. While our current diffusion purification strategy operates on a single-scale basis, we believe that incorporating multiscale mechanisms—either via multi-resolution feature fusion or hierarchical denoising—could further enhance robustness, especially in the presence of localized lesions or fine-grained aliasing artifacts. We leave this direction for future investigation.}






\appendix

\section{Proof of Theorem~1}

\noindent \textbf{Proof of Theorem~1:} For the first part, we begin by establishing the results in \eqref{eqn: KL theorem}. Utilizing the VE-SDE formulation of DMs, the conditional distributions $p_{0t}$ and $q_{0t}$ are expressed as per the following equations \cite{song2020score}. 
\begin{subequations}\label{eqn: cond gausses}
    \begin{align}
        p_{0t}(\mathbf{z}(t)\mid \mathbf{z}) = \mathcal{N}(\mathbf{z}(t) ; \mathbf{z}, (\sigma^2(t)-\sigma^2(0))\mathbf{I})\:, \\
        q_{0t}(\mathbf{z}(t)\mid \mathbf{z}_\textrm{pert}) = \mathcal{N}(\mathbf{z}(t) ; \mathbf{z}_\textrm{pert}, (\sigma^2(t)-\sigma^2(0))\mathbf{I})\:.
    \end{align}
\end{subequations}
Notably, these two distributions have different means, but share the same covariance. Consequently, the $D_{\textrm{KL}}$ can be obtained as 
\begin{equation}\label{eqn: KL big eq}
\notag
\begin{aligned}
D_{\textrm{KL}}(p_{0t}\mid\mid q_{0t}) = \frac{1}{2} \bigg( \log\Big(\frac{\det(\sigma^2 \mathbf{I})}{\det(\sigma^2 \mathbf{I})}\Big) + \textrm{Tr}\Big((\sigma^2\mathbf{I})^{-1}(\sigma^2\mathbf{I})\Big) \\+ (\mathbf{z}_\textrm{pert} - \mathbf{z})^T (\sigma^2\mathbf{I})^{-1} (\mathbf{z}_\textrm{pert} - \mathbf{z}) -n  \bigg) \:,
\end{aligned}
\end{equation}
where $\det(\cdot)$ (resp. $\textrm{Tr}(\cdot)$) denotes the determinant (resp. trace) of a matrix, and $\sigma^2 = \sigma^2(t)-\sigma^2(0)$ is used for brevity. Since $\log(1)=0$, the first term is zero. Given the definition of the trace and the identity matrix properties, the second term reduces to $n$ and cancels the last term. Since $\mathbf{A}^H \bm{\delta} = \mathbf{z}_\textrm{pert} - \mathbf{z}$, and $ (\mathbf{A}^H \bm{\delta})^T \mathbf{A}^H \bm{\delta} \geq 0$, then Equation \eqref{eqn: KL theorem} holds.

Subsequently, the numerator in \eqref{eqn: KL theorem} is more than or equal to $0$ (can only be zero if $\bm{\delta} = 0$), and is not a function of $t$. Moreover, since $\sigma(t) = \sigma_l(\sigma_u / \sigma_l)^t$, where $\sigma_l\in (0,1)$ and $\sigma_u >1$ are constants, it is evident that the denominator monotonically increases as $t$ increases.  

In conclusion, the rate of change of $D_{\textrm{KL}}(p_{0t}\mid\mid q_{0t})$ w.r.t. $t$ (as long as $\bm{\delta}\neq 0$) is less than $0$. Given the derivative of $\sigma(t)$ w.r.t. $t$ is $\frac{d\sigma(t)}{dt} = \sigma_l \log(\sigma_u / \sigma_l) (\sigma_u / \sigma_l)^t$, this is supported by
\begin{equation}\label{eqn: derivative KL}
\notag
\begin{aligned}
\frac{dD_{\textrm{KL}}(p_{0t}\mid\mid q_{0t})}{dt} = \frac{-\|\mathbf{A}^H\bm{\delta}\|^2 \sigma_l \log(\sigma_u/\sigma_l) (\sigma_u/\sigma_l)^{2t}}{\big( \sigma^2(t) - \sigma^2_l \big)^2}<0 \:.
\end{aligned}
\end{equation}
This inequality establishes that $D_{\textrm{KL}}(p_{0t}\mid\mid q_{0t})$ monotonically decreases as time travels from $t=0$ to $t=1$ while employing the forward process defined in \eqref{eqn: SCORE-BASED forward}. Consequently, the proof of the first part is complete.

The proof of the second part follows from \cite{song2020score} and \cite{pmlr-v162-nie22a}. Using the Fokker-Planck-Kolmogorov representation \cite{sarkka2019applied} for the forward process in \eqref{eqn: SCORE-BASED forward}, we write
\begin{subequations}\label{eqn: fokker planck}
    \begin{align}
        \frac{dp_{t}(\mathbf{z})}{dt} = \frac{1}{2}\nabla_{\mathbf{z}} \cdot \big( p_t(\mathbf{z}) \frac{d\sigma^2(t)}{dt} \nabla_\mathbf{z} \log p_t(\mathbf{z}) \big)\:, \\
        \frac{dq_{t}(\mathbf{z})}{dt} = \frac{1}{2}\nabla_{\mathbf{z}} \cdot \big( q_t(\mathbf{z}) \frac{d\sigma^2(t)}{dt} \nabla_\mathbf{z} \log q_t(\mathbf{z}) \big)\:.
    \end{align}
\end{subequations}
%

Employing the definition of the KL divergence, Equation~\eqref{eqn: fokker planck}, integration by parts, and assuming the smoothness and fast decay of $p_t(\mathbf{z})$ and $q_t(\mathbf{z})$, we can derive the derivative of the KL divergence w.r.t. $t$:
\begin{equation}\label{eqn: derivative KL 2}
\begin{aligned}
\frac{dD_{\textrm{KL}}(p_{t}\mid\mid q_{t})}{dt} = -\frac{1}{2} \frac{d\sigma^2(t)}{dt} D_\textrm{F}(p_t\mid\mid q_t) \leq 0 \:,
\end{aligned}
\end{equation}
%
%
where
\begin{equation}\label{eqn: Fisher}
\notag
\begin{aligned}
D_\textrm{F}(p_t\mid\mid q_t) = \int p_t(\mathbf{z}) \|\nabla_\mathbf{z} \log p_t(\mathbf{z}) - \nabla_\mathbf{z} \log q_t(\mathbf{z}) \|^2 d\mathbf{z} \geq 0 \:,
\end{aligned}
\end{equation}
%
%
denotes the Fisher divergence. Given that $\frac{d\sigma^2(t)}{dt} > 0$, the proof of the second part is thereby established.

\bibliographystyle{IEEEbib}
\bibliography{reference}
\begin{IEEEbiography}[{\includegraphics[width=1in,height=1.25in,clip,keepaspectratio]{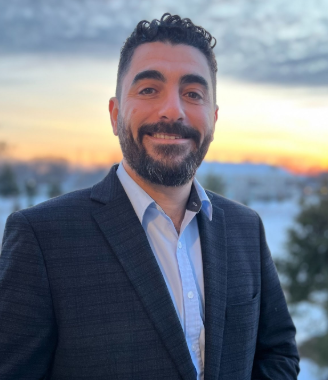}}]{Ismail R. Alkhouri}(Member, IEEE) is a Research Scientist at SPA Inc., providing technical support to the Information Innovative Office at the Defense Advanced Research Projects Agency (DARPA). He is also a Research Scholar at the University of Michigan (Electrical Engineering and Computer Science (EECS) Department) and Michigan State University (Computational Mathematics, Science, and Engineering (CMSE) Department). He received a Ph.D. in Electrical and Computer Engineering from the University of Central Florida in May 2023. From 2019 to 2022, he was a research intern at the Air Force Research Laboratory (Information directorate), and from July 2023 to December 2024, he was a Postdoctoral Researcher at Michigan State University (CMSE Department) and the University of Michigan (EECS Department). He is a recipient of the Rising Stars Award at the 2025 Conference on Parsimony and Learning (CPAL). He is also a recipient of the Outstanding Alumni Research Award at the 2025 MSU's CMSE 10th Anniversary workshop. His research focuses on computational imaging with deep generative models and differentiable methods for combinatorial optimization. His work was recognized as a finalist for best paper awards at ICASSP 2021 and MLSP 2023.
\end{IEEEbiography}

\begin{IEEEbiography}[{\includegraphics[width=1in,height=1.25in,clip,keepaspectratio]{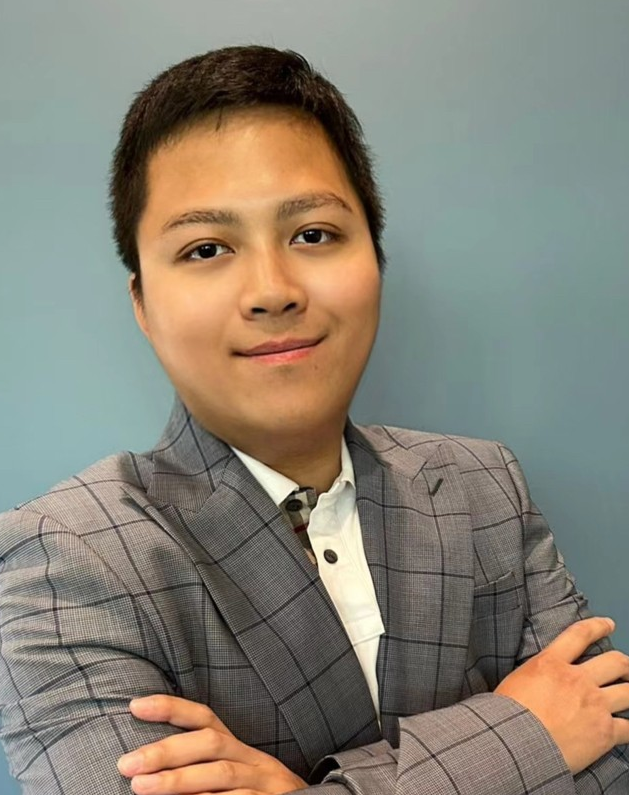}}]{Shijun Liang}(Member, IEEE)  received his B.S. degree in Biochemistry from the University of California, Davis, CA, USA, in 2017. In 2025, he received Ph.D. student in the Department of Biomedical Engineering at Michigan State University, East Lansing, MI, USA. His research focuses on machine learning and optimization techniques for solving inverse problems in imaging. Specifically, he is interested in machine learning based image reconstruction and in enhancing the robustness of learning-based reconstruction algorithms.
\end{IEEEbiography}

\begin{IEEEbiography}[{\includegraphics[width=0.95\textwidth]{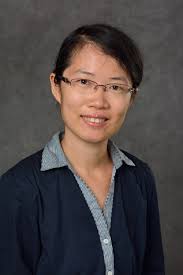}}]{Rongrong Wang}holds a B.S. in Mathematics and a B.A. in Economics from Peking University, China, as well as a Ph.D. in Applied Mathematics from the University of Maryland, College Park. She is currently an Associate Professor in both the Department of Computational Mathematics, Science and Engineering and the Department of Mathematics at Michigan State University. Prior to joining MSU, she was a postdoctoral researcher in the Department of Mathematics and the Department of Earth, Ocean, and Atmospheric Sciences at the University of British Columbia. Her research focuses on modeling and optimization techniques for extracting insights from data through computation. She specializes in the development of learning algorithms, optimization formulations, and numerical methods that offer theoretical performance guarantees and are scalable to large datasets. Her work has key applications in signal processing, machine learning, and inverse problems.
\end{IEEEbiography}

\begin{IEEEbiography}
[{\includegraphics[width=\textwidth]{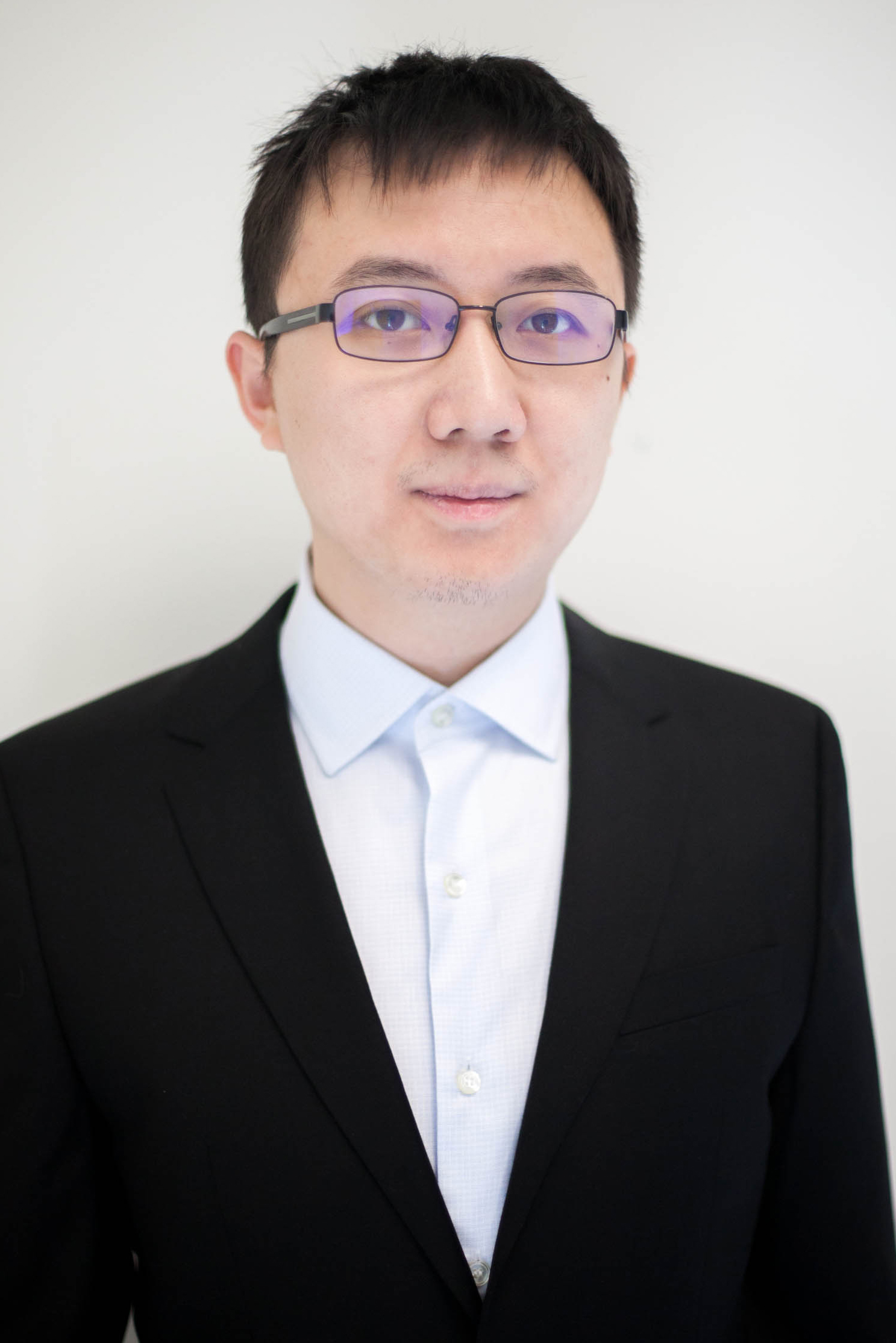}}]{Qing Qu}
    received his B.E. degree from Tsinghua University, Beijing, China, in 2011, and obtained his Ph.D. degree from Columbia University with Prof. John Wright in 2018. He was a Moore-Sloan fellow at NYU Center for Data Science from 2018 to 2020. His work has been recognized by multiple awards, including a Microsoft Ph.D. Fellowship in machine learning, an NSF Career Award in 2022, and an Amazon AWS AI Award in 2023. He is one of the founding organizers of the Conference on Parsimony and Learning (CPAL).
\end{IEEEbiography}

\begin{IEEEbiography}
[{\includegraphics[width=0.95\textwidth]{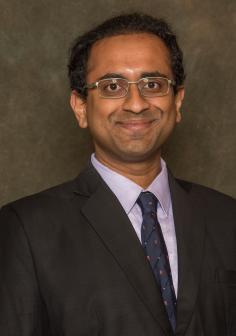}}]{Saiprasad Ravishankar} (Senior Member, IEEE) received the B.Tech. degree in Electrical Engineering
from the Indian Institute of Technology Madras,
Chennai, India, in 2008, and the M.S. and Ph.D. degrees in Electrical and Computer Engineering from the
University of Illinois at Urbana-Champaign, Urbana, IL, USA, in 2010 and 2014, respectively. He is currently an Assistant Professor with
the Departments of Computational Mathematics, Science and Engineering, and Biomedical Engineering,
Michigan State University, Michigan, USA. He was
an Adjunct Lecturer and a Postdoctoral Research Associate with the University
of Illinois at Urbana-Champaign from February to August, 2015. Since August 2015, he was a Postdoc with the
Department of Electrical Engineering and Computer Science at the University
of Michigan, Ann Arbor, MI, USA, and then a Postdoc Research Associate with the Theoretical
Division at Los Alamos National Laboratory, Los Alamos, NM, USA, from
August 2018 to February 2019. His research interests include signal and image
processing, biomedical and computational imaging, machine learning, inverse
problems, and large-scale data processing and optimization. He was the recipient
of the IEEE Signal Processing Society Young Author Best Paper Award in 2016. A paper he co-authored won a
Best Student Paper Award at the IEEE International Symposium on Biomedical
Imaging (ISBI) 2018 and other papers were award
finalists at the IEEE International Workshop on Machine Learning for Signal
Processing (MLSP) 2017, ISBI 2020, and Optica Imaging Congress, 2023. He is currently a member of the IEEE Machine Learning for Signal Processing Technical Committee. He has organized several special
sessions or workshops on computational imaging and machine learning themes
including at the Institute for Mathematics and its Applications (IMA), IEEE
Image, Video, and Multidimensional Signal Processing (IVMSP) Workshop
2016, MLSP 2017, ISBI 2018, the International Conference on Computer
Vision (ICCV) 2019 and 2021, etc.
\end{IEEEbiography}


\end{document}